\documentclass[12pt,epsf]{article}

\usepackage[dvips]{graphicx}
\usepackage{color}

\newcommand{\eeq}{\end{equation}}
\newcommand{\bea}{\begin{eqnarray}}
\newcommand{\eea}{\end{eqnarray}}
\newcommand{\AmSLaTeXe}{%
 $\mathcal A$\lower.4ex\hbox{$\!\mathcal M\!$}$\mathcal S$-\LaTeXe}

\usepackage{bm}

\usepackage{caption}
\begin{document}
\thispagestyle{empty}
\vspace*{-15mm}
{\bf OCHA-PP-348}\\

\vspace{15mm}
\begin{center}
{\Large\bf
Generalized Heisenberg-Euler formula in Abelian gauge theory with parity violation\\
}
\vspace{7mm}

\baselineskip 18pt
{\bf Kimiko Yamashita$^{1, 2}$, Xing Fan${}^{3, 4}$, Shusei Kamioka${}^{3}$, \\
Shoji Asai${}^{3}$, and Akio Sugamoto${}^{1, 5}$
}
\vspace{2mm}

{\it
${}^{1}$Department of Physics, Graduate School of Humanities and Sciences,\\
Ochanomizu University, 2-1-1 Ohtsuka, Bunkyo-ku, Tokyo 112-8610, Japan\\
${}^{2}$Program for Leading Graduate Schools,\\
Ochanomizu University, 2-1-1 Ohtsuka, Bunkyo-ku, Tokyo 112-8610, Japan\\
${}^{3}$Department of Physics, Graduate School of Science, The University of Tokyo, 7-3-1 Hongo, Bunkyo-ku, Tokyo 113-0033, Japan\\
${}^{4}$Department of Physics, Harvard University, Cambridge, MA02138, USA\\
${}^{5}$Tokyo Bunkyo SC, the Open University of Japan, Tokyo 112-0012, Japan \\
}

\vspace{10mm}
\end{center}
\begin{center}
\begin{minipage}{14cm}
\baselineskip 16pt
\noindent
\begin{abstract}
A generalized Heisenberg-Euler formula is given for an Abelian gauge theory having vector as well as axial vector couplings to a massive fermion. So, the formula is applicable to a parity-violating theory.  The gauge group is chosen to be $U(1)$.  The formula is quite similar to that in quantum electrodynamics, but there is a complexity in which one factor (related to spin) is expressed in terms of the expectation value. The expectation value is evaluated by the contraction with the one-dimensional propagator in a given background field. The formula affords a basis to the vacuum magnetic birefringence experiment, which aims to probe the dark sector, where the interactions of the light fermions with the gauge fields are not necessarily parity conserving.
\end{abstract}

\end{minipage}
\end{center}

\baselineskip 18pt
\def\thefootnote{\fnsymbol{footnote}}
\setcounter{footnote}{0}

\newpage

\section{Introduction}
The low-energy effective Lagrangian $\mathcal{L}^{\mathrm{QED}}_{\mathrm{eff}}$ of quantum electrodynamics (QED) was derived in 1936 by W. Heisenberg and H. Euler; the corresponding formula is called the Heisenberg-Euler formula \cite{Heisenberg-Euler} (see also the review articles \cite{review of H-E1,review of H-E2,review of H-E3}), which reads:
\begin{eqnarray}
\mathcal{L}^{\mathrm{QED}}_{\mathrm{eff}}=-\mathcal{F} +\frac{8}{45}\left(\frac{\alpha^2}{m_e^4}\right)\mathcal{F}^2 + \frac{14}{45}\left(\frac{\alpha^2}{m_e^4}\right)\mathcal{G}^2+ \cdots,  \label{H-E}
\end{eqnarray}
where
\begin{eqnarray}
\mathcal{F} =\frac{1}{4}F_{\mu\nu}F^{\mu\nu}=\frac{1}{2}\left(\bm{B}^2-\bm{E}^2\right), ~\mathcal{G} =\frac{1}{4}F_{\mu\nu} \tilde{F}^{\mu\nu}=\bm{E} \cdot \bm{B}.  \label{F and G}
\end{eqnarray}
Here, $\bm{E}$ and $\bm{B}$ are electric and magnetic fields of QED, respectively, $\alpha=\frac{e^2}{4\pi}$ is the fine-structure constant, and $m_e$ is the electron mass. The dual field strength is defined by $\tilde{F}_{\mu\nu}=\frac{1}{2}\epsilon_{\mu\nu\lambda\rho}F^{\lambda\rho}$, where $\epsilon_{\mu\nu\lambda\rho}$ is the totally antisymmetric tensor with $\epsilon_{0123}=1$.  The first term of the formula is the usual QED action for the photon, but the next term of $O(\alpha^2/m_e^4)$ gives a nonlinear equation of motion for the photon field.

This nonlinear term gives the interesting effect of ``vacuum magnetic birefringence'', which is seen in experiments in which the change of the polarization of a laser beam is measured after passing through external strong magnetic fields--see~\cite{BMV,PVLAS}, and the OVAL (Observing VAcuum with Laser) experiment~\cite{OVAL1,OVAL2,OVAL3}.

The Heisenberg-Euler (H-E) formula is only applicable to QED.  If we intend to search the dark sector (the sector coupled extremely weakly to the real world) with such experiments, a more general formula is necessary, that is applicable to a theory in which the gauge field couples to the fermions with vector as well as axial vector couplings.  An example of a consistent theory is the standard model (SM) built based on the $SU(3)_C \times SU(2)_L \times U(1)_Y$ gauge theories, where the general-type interactions with vector as well as axial vector couplings to fermions appear without any problem.  Parity is always violated in the weak interactions by the axial vector coupling.

So, we will derive in this paper a generalized H-E formula in a more general case than QED, having the interaction of vector as well as axial vector couplings with the fermion.  For simplicity we will study the Abelian case in this paper.  

Physical implications obtained from the generalized H-E formula, and how to observe its physical effects via the vacuum magnetic birefringence experiments are discussed in a separate paper~\cite{application to OVAL}.  The more general case with non-Abelian gauge symmetries will be studied in the near future.

\section{Generalized Heisenberg-Euler formula} 
In this section we will give the generalized H-E formula for a model in which a fermion field $\psi(x)$ couples to a gauge field $A_{\mu}(x)$ in a general way with a vector coupling $g_V$ as well as an axial vector coupling $g_A$.  The action of this model is
\begin{eqnarray}
S_{\psi}(m)=\int d^4 x~ \bar{\psi}(x) \left[\gamma^{\mu} \left( i\partial_{\mu} -(g_V+g_A \gamma_5) A_{\mu}\right) - m \right] \psi(x), \label{original action}
\end{eqnarray}
which gives the effective action $S_{\mathrm{eff}}[A_{\mu}]$ and the effective Lagrangian $\mathcal{L}_{\mathrm{eff}}[A_{\mu}]$ for a given background field configuration $A_{\mu}(x)$ as 
\begin{eqnarray}
 S_{\mathrm{eff}}[A_{\mu}] &=&\int d^4x~\mathcal{L}_{\mathrm{eff}}[A_{\mu}]= -i \ln \left[ \int \mathcal{D} \psi(x) \mathcal{D} \bar{\psi}(x) e^{iS_{\psi}(m)} \right] \nonumber \\
&=&-i \mathrm{Tr} \ln \left[\gamma^{\mu} \left( i\partial_{\mu} -(g_V+g_A \gamma_5) A_{\mu}\right) - m \right].
 \end{eqnarray}

In deriving the generalized H-E formula, or the effective Lagrangian $\mathcal{L}_{\mathrm{eff}}$ given above, we follow faithfully a beautiful method developed by J. Schwinger \cite{Schwinger} in QED.  He uses the proper time formalism developed in Refs.~\cite{Fock and Nambu1, Fock and Nambu2}.  Our contribution is only to simplify his formalism a little with the help of the path integral method \cite{Feynman}.  In deriving the effective action properly renormalization of the divergences is necessary, so the theory should be anomaly free to guarantee renormalizability.  However, what we are going to study is the simplified model in Eq.~(\ref{original action}), which may not be free from anomalies.  It is true that we will finally apply the derived formula in this paper to the anomaly free theory, by adding other fields so as to cancel anomalies. However, the formula derived in this paper can give a contribution of one species of particle in the effective action of the anomaly-free theory.\footnote{The effective action is a sum of contributions from the different particle species, depending on their mass such as $\mathcal{L}_{\mathrm{eff}}= \sum_{i} \frac{1}{m_i^4} (a_i \mathcal{F}^2+ b_i \mathcal{G}^2+i c_i \mathcal{F}\mathcal{G}).$  On the other hand, the anomaly cancellation condition is given by the representation and the number of particle species, and it does not depend on their masses.  Since the effective action is dominated by the particle with the minimum mass, we keep only this particle in the effective action, but understand that there exist other heavier particles which cancel anomalies with the minimum mass particle, even if the heavier particles disappear from the effective action.} 

In this respect we have to comment on the excellence of Schwinger's proper time formalism, which is that it is applicable even to anomalous theories.  This can be understood from the fact that the correct anomalous axial gauge Ward identity is derived in the same article \cite{Schwinger} in 1951.  So we will follow his method faithfully. 

Now, we have to make a chiral transformation $\psi'=i\gamma_5 \psi$ to obtain another action $S_{\psi}(m)=S_{\psi'}(-m)$, where the sign of the fermion mass is reversed.  Related to the chiral rotation the Jacobian $\mathcal{J}$ appears\footnote{The importance of the Jacobian in the anomalous theory is pointed out in \cite{Fujikawa1, Fujikawa2}.} This $\mathcal{J}$ is, however,  shown to be 1 in Appendix B, so we may forget about the Jacobian in the following sections.

For a given background field $A_{\mu}(x)$, the one-loop effective action $S_{\mathrm{eff}}(A)$ reads 
\begin{eqnarray}
 e^{iS_{\mathrm{eff}}(A)}  \equiv  Z_{\psi}(A) &=& \int \mathcal{D} \psi(x) \mathcal{D} \bar{\psi}(x) e^{iS_{\psi}(m)}= \int \mathcal{D} \psi'(x) \mathcal{D} \bar{\psi}'(x) e^{iS_{\psi'}(-m)} \nonumber \\
&=& \left(\int \mathcal{D} \psi(x) \mathcal{D} \bar{\psi}(x) e^{iS_{\psi}(m)} \times  \int \mathcal{D} \psi'(x) \mathcal{D} \bar{\psi}'(x) e^{iS_{\psi'}(-m)}\right)^{\frac{1}{2}}.~~~~
 \end{eqnarray}
After the path integration, we obtain\footnote{$(\mathrm{det}\hat{O})^n=e^{n\mathrm{Tr} \ln \hat{O}}$.}
\begin{eqnarray}
&&S_{\mathrm{eff}}(A)  \\
&=&(-i) \frac{1}{2} \mathrm{Tr} \ln \left[- \left\{ \gamma^{\mu} \left( i\partial_{\mu} -(g_V+g_A \gamma_5) A_{\mu}\right) - m \right\} \cdot \left\{ \gamma^{\nu} \left( i\partial_{\nu} -(g_V+g_A \gamma_5) A_{\nu}\right) + m \right\} \right] \nonumber \\
&+& i \frac{1}{2} \ln \mathcal{J} = (-i) \frac{1}{2} \mathrm{Tr} \ln (\hat{H}+ m^2),
\end{eqnarray}
where the quantum Hamiltonian $\hat{H}$ is given by
\begin{eqnarray}
\hat{H}=-\left\{ \gamma^{\mu} \left( i\partial_{\mu} -(g_V+g_A \gamma_5) A_{\mu}\right) \right\}^2= \left\{ \gamma^{\mu} D_{\mu}(A(x))\right\}^2=\{\rlap{\it{D}}\;/~(A)\}^2,
\end{eqnarray}
with the covariant derivative $D_{\mu}(A(x)) \equiv \partial_{\mu} +i(g_V+g_A \gamma_5) A_{\mu}$ and $\rlap{\it{D}}~/(A)=\gamma^{\mu} D_{\mu}(A)$. To introduce the Hamiltonian, $\hat{H}+m^2$ is to be positive after Wick rotation.

This Hamiltonian $\hat{H}$ becomes 
\begin{eqnarray}
&&\hat{H} =-\left( i \partial_{\mu} - g_V A_{\mu}(x)\right)^2 +\left(g_A A_{\mu}(x)\right)^2 + \frac{1}{2}(g_V+g_A \gamma_5) \sigma^{\mu\nu} F_{\mu\nu}(x) \label{Hamiltonian 1}  \nonumber \\
&&~~~~~~+i \sigma^{\mu\nu}g_A \gamma_5(A_{\mu}i \partial_{\nu}-A_{\nu}i \partial_{\mu} )\\
&&=-\left(i \partial_{\mu} - g_V\frac{1}{2}x^{\nu}F_{\nu\mu}(x)\right)^2 -\frac{1}{4}x^{\mu}\left(g_A^2  F_{\mu\lambda}F^{\lambda\nu}\right) x_{\nu}  \nonumber \\
&&~~~+\frac{1}{2}(g_V+g_A \gamma_5) \sigma^{\mu\nu} F_{\mu\nu} +i \frac{1}{2} \sigma^{\mu\nu}g_A \gamma_5(x^{\lambda}F_{\lambda\mu}\; i \partial_{\nu}-x^{\lambda}F_{\lambda\nu}\; i \partial_{\mu}),
 \label{Hamiltonian 2}
\end{eqnarray}
where $\gamma^{\mu}\gamma^{\nu}=g^{\mu\nu} - i \sigma^{\mu\nu}$ has been used with the definition of $\sigma^{\mu\nu}=\frac{i}{2} [\gamma^{\mu}, \gamma^{\nu}]$ in the notation of $\gamma$ matrices by Bjorken and Drell.  
In the above calculation, since we are going to derive the low-energy effective action {\it \`a la}  Heisenberg and Euler, we have assumed that the field strength $F_{\mu\nu}=\partial_{\mu} A_{\nu}(x)-\partial_{\nu} A_{\mu}(x)$ is constant, and the gauge field has been expressed by $A_{\mu}(x)=\frac{1}{2} x^{\lambda} F_{\lambda\mu}$. The gauge condition $\partial_{\mu} A^{\mu}=0$ is used.  In the expressions (\ref{Hamiltonian 1}), the last term proportional to $\sigma^{\mu\nu}(A_{\mu}i \partial_{\nu}-A_{\nu}i \partial_{\mu})$ remains when $g_A \ne 0$, with which the dynamics of $x^{\mu} $ and the dynamics of the spin $\sigma^{\mu\nu}$ can't be separated as in the vector-like theory of QED.  This gives the complexity in the parity-violating theory not existing in a party-conserving theory such as QED.

Here, it is useful to understand that in the case of a constant external magnetic field in the z direction, the vector coupling ($g_V$) gives a cyclotron motion in the $(x, y)$ plane, while the axial vector coupling ($g_A$) gives a harmonic oscillator in the $(x, y)$ plane, which is not a simple harmonic oscillator but is modified by the interaction with the spin.  The Landau level of the cyclotron motion in the former case can be seen as a one-dimensional harmonic oscillator with a canonical pair $(p_x, p_y)$, while the latter modified harmonic oscillator is two dimensional with two canonical pairs, s$(x, p_x)$ and $(y, p_y)$.  This difference of dimensionality becomes manifest in the final expression of the effective action.


If we use the following identity (shown by differenting both sides by $\hat{H}$), we have
\begin{eqnarray}
-\frac{i}{2}~\mathrm{Tr} \ln (\hat{H}+m^2)=\frac{i}{2} \int_0^{\infty}~ \frac{ds}{s}~\mathrm{Tr} e^{-i (\hat{H}+m^2) s},
\end{eqnarray}
and the effective action becomes the following simple expression,
\begin{eqnarray}
S_{\mathrm{eff}}(A)= \frac{i}{2} \int_0^{\infty}~ \frac{ds}{s}~e^{-i m^2s}~\mathrm{Tr }(e^{-i \hat{H} s}),
\end{eqnarray}
where the Hamiltonian is given by Eq.~(\ref{Hamiltonian 1}).  So, the problem is reduced to the quantum mechanics of a point particle in which the position of the particle at a proper time $s$ is described by $x^{\mu}(s)$, and its Hamiltonian is given by $\hat{H}$.  The operator $e^{-i \hat{H} s}$ is the shift operator for the time $s$ which is the ``proper time" studied in Refs.~\cite{Fock and Nambu1, Fock and Nambu2}.  There are two ways to estimate $\mathrm{Tr} e^{-i \hat{H} s}$, the operator formalism or the path integral method.  


Also there are a number of ways to estimate this effective action.

Here we will proceed with a ``hybrid method.''  That is, we consider the position variables $(x^{\mu}, p^{\mu})$ to be ``classical'' ones and apply the path integral quantization.  On the other hand, the spin matrices are preserved in the initial form.  The spin matrices change the direction of the ``spin'' (more precisely, the direction of the four-component spinor), so that the spin matrices can be considered as if they were the ``quantum operators'' acting on the ``spin.''  In this sense we call this method as  the hybrid one.  Another way is to consider the spin also as purely classical variables and apply the path integral quantization to them \cite{Bern-Kosower1, Bern-Kosower2, Strassler}.  In the latter way we lift the gamma matrices $\gamma^{\mu}$ to the anticommuting Ramond field $\psi^{\mu}(s)$ \cite{Ramond1, Ramond2}   
\begin{eqnarray}
\frac{1}{\sqrt{2}} \gamma^{\mu} \to \psi^{\mu}(s).
\end{eqnarray}
which represents the spin at $s$, and is required to satisfy the anticommutation relation
\begin{eqnarray}
\{ \psi^{\mu}, ~\psi^{\nu} \}=g^{\mu\nu},
\end{eqnarray}
to be consistent with the Clifford algebra $\{\gamma^{\mu}, ~\gamma^{\nu}\}=2 g^{\mu\nu}$.  The former and the latter methods are equivalent.  In our problem, taking the latter method does not seem to make the calculation simpler than the former method, and so we will adopt the former method.  This is the method adopted by Schwinger and it works well even in a parity-violating theory. 

Now, in the hybrid method, the ``classical'' (or ``hybrid'') Hamiltonian $\tilde{H}$ reads
\begin{eqnarray}
&&\tilde{H} =-\left( p_{\mu} - g_V A_{\mu}(x)\right)^2 +\left(g_A A_{\mu}(x)\right)^2 + \frac{1}{2}(g_V+g_A \gamma_5) \sigma^{\mu\nu} F_{\mu\nu}(x)\nonumber \\
&&~~~~~~+i \sigma^{\mu\nu}g_A \gamma_5(A_{\mu}p_{\nu}-A_{\nu}p_{\mu} )\\
&&=-\left(p_{\mu} - g_V\frac{1}{2}x^{\nu}F_{\nu\mu}(x)\right)^2 -\frac{1}{4}x^{\mu}\left(g_A^2  F_{\mu\lambda}F^{\lambda\nu}\right) x_{\nu}  \nonumber \\
&&~~~+\frac{1}{2}(g_V+g_A \gamma_5) \sigma^{\mu\nu} F_{\mu\nu} +i \frac{1}{2} \sigma^{\mu\nu}g_A \gamma_5(x^{\lambda}F_{\lambda\mu}\; p_{\nu}-x^{\lambda}F_{\lambda\nu}\; p_{\mu}),
 \label{Hybrid Hamiltonian}
\end{eqnarray}

The corresponding hybrid Lagrangian $\tilde{L}$ can be obtained as follows:
\begin{eqnarray}
&&\tilde{L}(x^{\mu}(s), \dot{x}^{\mu}(s)) = -\frac{1}{4}(\dot{x}^{\mu})^2 +A_{\mu} \left(g_V \; g^{\mu\nu}+i g_A\; \gamma_5 \sigma^{\mu\nu} \right) \dot{x}_{\nu} +2 g_A^2 (A_{\mu})^2 \nonumber \\
&-&\frac{1}{2} \sigma^{\mu\nu} (g_V+g_A \gamma_5) F_{\mu\nu} \\
&=&-\frac{1}{4}(\dot{x}^{\mu})^2 +\frac{1}{2} x^{\mu} F_{\mu\lambda} \left(g_V \; g^{\lambda\nu}+i g_A\; \gamma_5 \sigma^{\lambda\nu} \right) \dot{x}_{\nu} -\frac{1}{2} g_A^2~ x^{\mu} F_{\mu\lambda}F^{\lambda\nu} x_{\nu} \nonumber \\
&-&\frac{1}{2} \sigma^{\mu\nu} (g_V+g_A \gamma_5) F_{\mu\nu}
\label{Hybrid Lagrangian of point particle}
\end{eqnarray}
where the ``dot'' denotes the derivative with respect to $s$.  Using this hybrid Lagrangian, we have
\begin{eqnarray}
\mathrm{Tr}e^{-i \hat{H} s} &=&\int d^4 x ~\mathrm{tr}~ \langle x(s), a(s) | x(0), b(0) \rangle \\
&=& \int d^4 x ~\mathrm{tr}~\int_{x^{\mu}(0)=x^{\mu}}^{x^{\mu}(s)=x^{\mu}} \mathcal{D} x^{\mu}(s') ~ e^{i\int_0^s~ ds'~\tilde{L}\left(x(s'), ~\dot{x}(s') \right)},\label{hybrid transition amplitude}
\end{eqnarray}
where the initial and the final spin states are denoted by $b(0)$ and $a(s)$, respectively.  ``Tr'' consists of the trace of $x^{\mu}$, or $\int d^4 x$, as well as the trace of the four-component ``spin'' (spinor) degrees of freedom, the latter of which (trace of the spin matrices) is denoted by ``tr,'' following Schwinger. 

To estimate the transition amplitude Eq.~(\ref{hybrid transition amplitude}) it is useful to know the general structure of the action integral,
\begin{eqnarray}
\tilde{S}(s)=\int_0^s ds' \tilde{L}(s') = \int_0^s ds' \left( A(s') + \frac{1}{2} B_{\mu\nu} (s') \sigma^{\mu\nu} \right), 
\end{eqnarray}
where
\begin{eqnarray}
&&A(s')=-\frac{1}{4}(\dot{x}^{\mu})^2 +\frac{1}{2} g_V x^{\mu} ( F_{\mu\nu})\dot{x}^{\nu} - \frac{1}{2} g_A^2 x^{\mu} (F_{\mu\lambda}F^{\lambda\nu})x_{\nu},  \label{A(s')} \\
&&B_{\mu\nu}(s')= g_A \frac{1}{2} \epsilon_{\mu\nu\beta\gamma} x_{\alpha}F^{\alpha\beta}\dot{x}^{\gamma} -(g_V F_{\mu\nu} -i g_A \tilde{F}_{\mu\nu}), \label{B(s')}
\end{eqnarray}
and $\gamma_5 \sigma_{\mu\nu}= -\frac{i}{2} \epsilon_{\mu\nu\alpha\beta} \sigma^{\alpha\beta}$ has been used.   Defining the integrals 
\begin{eqnarray}
\bar{A}(s)=  \int_0^s ds' A(s'), ~\bar{B}_{\mu\nu}(s)=\int_0^s ds' B_{\mu\nu}(s'),
\end{eqnarray}
the transition amplitude becomes
\begin{eqnarray}
&&\mathrm{tr} \langle x(s), a(s) | x(0), b(0) \rangle =\mathrm{tr} \int \mathcal{D} x^{\mu}(s') ~ e^{i\tilde{S}(s)}  \\
&&= \mathrm{tr} \int \mathcal{D} x^{\mu}(s') ~ e^{i\bar{A}(s)} e^{i\frac{1}{2}\bar{B}_{\mu\nu}(s) \sigma^{\mu\nu}}.
\end{eqnarray}
The second factor can be estimated, by using $\frac{1}{2}\{\sigma^{\mu\nu}, \sigma^{\lambda\rho}\}=g^{\mu\lambda}g^{\nu\rho}-g^{\mu\rho}g^{\nu\lambda}- i \epsilon^{\mu\nu\lambda\rho} \gamma_5$, as follows:
\begin{eqnarray}
&&\mathrm{tr} \langle x(s), a(s) | x(0), b(0) \rangle = \mathrm{tr} \int \mathcal{D} x^{\mu}(s') ~ e^{i\bar{A}(s)} \nonumber \\
&& \times \left(\cos \sqrt{2\bm{\mathcal{H'}}(s)} +\frac{i}{2} \bar{B}_{\mu\nu}(s) \sigma^{\mu\nu} \frac{\sin \sqrt{2\bm{\mathcal{H'}}(s)}}{\sqrt{2\bm{\mathcal{H'}}(s)}}  \right),
\end{eqnarray}
where
\begin{eqnarray}
&&\bm{\mathcal{H'}}(s) =  \bar{\bm{\mathcal{F'}}}(s) - i \gamma_5 \bar{\bm{\mathcal{G'}}}(s), \\
&&\bar{\bm{\mathcal{F'}}}(s)= \frac{1}{4} \bar{B}_{\mu\nu}(s)\bar{B}^{\mu\nu}(s), ~~
\bar{\bm{\mathcal{G'}}}(s)= \frac{1}{4} \bar{B}_{\mu\nu}(s)\bar{\tilde{B}}^{\mu\nu}(s). \label{F and G by B(s')}
\end{eqnarray}

After taking the ``tr'' for the spinor indices, the $\sigma^{\mu\nu}$-dependent terms vanish and we have
\begin{eqnarray}
&&\mathrm{tr} \langle x(s), a(s) | x(0), b(0) \rangle \nonumber \\
&&=\int \mathcal{D} x^{\mu}(s') ~ e^{i\bar{A}(s)} \times 2 \left(\cos \sqrt{2(\bar{\bm{\mathcal{F'}}}(s)+i \bar{\bm{\mathcal{G'}}}(s))} +\cos \sqrt{2(\bar{\bm{\mathcal{F'}}}(s)-i \bar{\bm{\mathcal{G'}}}(s))} \right), ~~~~~~\label{general transition amplitude}
\end{eqnarray}
where the contribution to ``tr'' is two for $\gamma_5=1$ (the right-handed spinor) and two for $\gamma_5=-1$ (the left-handed spinor).

This is the application of the technique of Schwinger \cite{Schwinger}  for summing the spin degree of freedom.
Therefore, if we introduce $\bar{\bm{X}}'_{\pm} (s) = \sqrt{2\left(\bar{\bm{\mathcal{F'}}}(s) \pm i \bar{\bm{\mathcal{G'}}}(s)\right)}$, then Eq.~(\ref{general transition amplitude}) can be written as
\begin{eqnarray}
\mathrm{tr} \langle x(s), a(s) | x(0), b(0) \rangle=\left\langle 2 \left(\cos \bar{\bm{X}}'_{+}(s) +\cos \bar{\bm{X}}'_{-}(s) \right) \right\rangle,
\end{eqnarray}
where the expectation value symbol $\langle \cdots \rangle$ means
\begin{eqnarray}
\langle \bar{O}(s) \rangle \equiv \int \mathcal{D} x^{\mu}(s') ~ e^{i\bar{A}(s)} \bar{O}(s).  \label{expectation value}
\end{eqnarray}

The explicit forms of $\bar{\bm{\mathcal{F'}}}(s)$ and $\bar{\bm{\mathcal{G'}}}(s)$ are given as follows:
\begin{eqnarray}
4\bar{\bm{\mathcal{F'}}}(s) &=& \int_0^s \int_0^s ds'  ds'' 
\frac{1}{2} g_A^2 \left[ x^{\alpha}(s') (F_{\alpha\gamma}F^{\gamma\beta}) x_{\beta}(s'')(\dot{x}^{\mu}(s')\dot{x}_{\mu}(s'')) \right. \nonumber \\
&&~~~~~~~~~~~~~~~~~~~~~~~\left. +(x^{\alpha}(s') F_{\alpha\beta} \dot{x}^{\beta}(s''))  (x^{\gamma}(s'') F_{\gamma\delta} \dot{x}^{\delta}(s'))  \right] \nonumber \\
&-&s \int_0^s ds' x^{\alpha}(s') 2g_A \left( g_V F_{\alpha\gamma} \tilde{F}^{\gamma\beta} +i g_A F_{\alpha\gamma}F^{\gamma\beta} \right) \dot{x}_{\beta}(s') \nonumber \\
&+& s^2 \left[(g_V^2+g_A^2) (F_{\mu\nu}F^{\mu\nu} ) -2i g_Vg_A (F_{\mu\nu}\tilde{F}^{\mu\nu} ) \right],  \label{F'(s)} \\
4\bar{\bm{\mathcal{G'}}}(s)&=& \int_0^s \int_0^s ds'  ds''  \left(-\frac{1}{2}g_A^2 \right) \epsilon^{\mu\nu\mu'\nu'} (x^{\alpha} F_{\alpha \mu} \dot{x}_{\nu} )(s')(x^{\alpha'} F_{\alpha' \mu'} \dot{x}_{\nu'} )(s'') \nonumber \\
&+& s \int_0^s ds'  
x^{\alpha}(s')2 g_A \left( g_V F_{\alpha\gamma} F^{\gamma\beta} - i g_A F_{\alpha\gamma}\tilde{F}^{\gamma\beta} \right) \dot{x}_{\beta}(s') \nonumber \\
&+& s^2 \left[ (g_V^2+g_A^2) (F_{\mu\nu}\tilde{F}^{\mu\nu} ) +2i g_Vg_A (F_{\mu\nu}F^{\mu\nu} ) \right]. \label{G'(s)}
\end{eqnarray}

Taking the limit $g_A \to 0$, or discarding the parity-violating interactions, we can reproduce the QED results,
\begin{eqnarray}
\bar{\bm{\mathcal{F'}}}(s) |_{g_A \to 0} = s^2 g_V^2~ \mathcal{F}, ~~\bar{\bm{\mathcal{G'}}}(s)|_{g_A \to 0} =s^2 g_V^2 ~\mathcal{G}, \\
\bar{\bm{X}}'_{+}=s g_V X, ~\mathrm{and}~ \bar{\bm{X}}'_{-}=s g_V X^{\dagger},
\end{eqnarray}
where $X=\sqrt{2(\mathcal{F}+i \mathcal{G})}$ is the quantity which plays the important role in the definition of the eigenvalues for the background field and in the H-E formula in QED--see Eqs.~(\ref{eigen-values1}) and (\ref{eigen-values2}) in the next section.


In QED $\bar{\bm{ \mathcal{F'}}}(s')$ and $\bar{\bm{\mathcal{G'}}}(s')$ don't depend on $x^{\mu}(s')$ and $\dot{x}^{\mu}(s')$, so Eq.~(\ref{general transition amplitude}) can be easily estimated without taking the expectation value.  Here, however, the spin-dependent factor $2(\cos \bar{\bm{X}}'_{+}(s)+\cos \bar{\bm{X}}'_{-}(s))$ is not independent of $x^{\alpha}$ and $\dot{x}_{\beta}$, but is nonlinear in terms of them for $g_A \ne 0$, so that the path integration over $\int \mathcal{D} x^{\mu}(s')$ becomes non-Gaussian, and perturbation theory is necessary for the parity-violating theory with $g_A \ne 0$.  This is the complexity that appears only in the parity-violating theory and not in QED.

The details of the perturbation theory will be discussed in the following sections.

Formally, the expectation value in Eq.~(\ref{expectation value}) is obtained in the perturbation theory as
\begin{eqnarray}
&&\mathrm{Tr} \langle x(s), a(s) | x(0), b(0) \rangle \nonumber \\
&&=\int d^4 x~ \langle x(s)| x(0)\rangle' \times \left\langle 2 \left(\cos \bar{\bm{X}}'_{+}(s) +\cos \bar{\bm{X}}'_{-} (s)\right) \right\rangle' \nonumber \\
&&=\int d^4x ~\langle x(s)| x(0)\rangle' \times \left\{ 2 \left(\cos \bar{\bm{X}}'_{+}(s)+\cos \bar{\bm{X}}'_{-}(s) \right) [x^\lambda(s') \to (-i)\delta/\delta j_{\lambda}(s')] \right.\nonumber \\
&&~~~~~~~~~~~~~~~~~~~~~~~~~\times \left. \left. e^{-i \int_0^s ds' \int_0^s ds'' \sum_{\alpha\beta} j^{\alpha}(s') \Delta(s'-s'')_{\alpha\beta} j^{\beta}(s'')} \right\} \right\vert_{j_{\lambda=0}},  \label{Feynman rule}
\end{eqnarray}
where the propagator $\Delta(s'-s'')_{\nu}^{~\lambda}$ is defined by
\begin{eqnarray}
\left\{ {\delta_{\mu}}^{\nu} \frac{\partial^2}{\partial s^{'2}} + 2{(g_V \bm{F} )_{\mu}}^{\nu}\frac{\partial}{\partial s^{'}} -2 {(g_A^2 \bm{F}^2 )_{\mu}}^{\nu} \right\}\Delta_{\nu}^{~\lambda}(s'-s'')=\delta_{\mu}^{~\lambda}\delta(s'-s''), \label{def of the propagator}
\end{eqnarray}
and its expression is derived in Appendix A.

Here, $\langle x(s)| x(0)\rangle'$ is the free part, that is, the spin-independent transition amplitude of the particle from the position $x^{\mu}$ to the same position $x^{\mu}$ with the unperturbed Lagrangian $A(s')$, namely
\begin{eqnarray}
\langle x(s)| x(0)\rangle'=\int^{x^{\mu}(s)=x^{\mu}}_{x^{\mu}(0)=x^{\mu}} \mathcal{D} x^{\mu}(s') ~ e^{i\int_0^s ds' A(s')},
\end{eqnarray}
while the second expectation value is denoted with ``prime''  and is the interaction part, after separating the free part $\langle x(s)| x(0)\rangle'$.

The original ``Tr'' consists of the trace on the position $x^{\mu}$ and that on the spin.  The former trace appears as $\int d^4 x$ with the constraint that the path is closed, $x^{\mu}(s)=x^{\mu}(0)=x^{\mu}$, while the latter traces ``tr'' are taken explicitly.  

Now the effective Lagrangian density reads
\begin{eqnarray}
\mathcal{L}_{\mathrm{eff}}(x)&=&\frac{i}{2} \int_0^{\infty}~ \frac{ds}{s}~e^{-i m^2s}~ \langle x(s) | x(0) \rangle^{\prime} \times \left\langle 2 \left(\cos \bar{\bm{X}}'_{+}(s) +\cos \bar{\bm{X}}'_{-} (s) \right) \right\rangle'.~~~ \label{effective Lagrangian 1}
\end{eqnarray}

The task remaining is to estimate $\langle x(s) | x(0) \rangle^{\prime}$ explicitly. So far we have identified $x^{\mu}$ to be the initial and final points of the path $\{x^{\mu}(s');~0 \le s' \le s\}$, but the more convenient way is to forget about the initial and final points, and to consider the closed path $C(x_0)$ with center of mass coordinates specified by $x^{\mu}_0$--see Fig.~1. This is a standard way to study the closed string.  Then, we can proceed as follows: 

\begin{eqnarray}
&&\int d^4x \langle x(s) | x(0) \rangle^{\prime}=\int d^4 x_0 \int_{C(x_0)} \mathcal{D} x^{\mu}(s')~ e^{i\int_0^s~ ds'~A\left(x(s'), ~\dot{x}(s')\right)} \\
&&\equiv \int d^4 x_0 ~\langle C(x_0); s \rangle^{\prime}, \label{transition amplitude 2}
\end{eqnarray}
where the prime indicates the spin-independent part.

\begin{figure}[ht]
\center
\includegraphics[width=4cm]{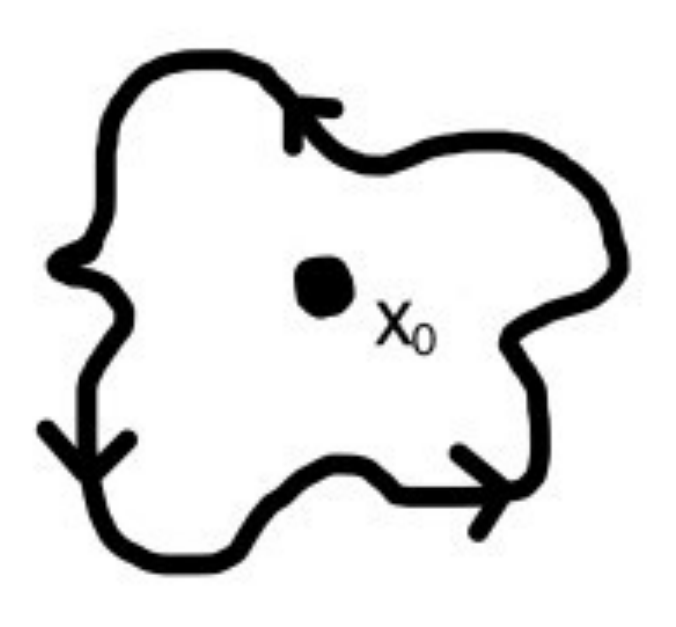}
\caption{A closed path $C(x_0)$ with center of mass coordinates specified by $x^{\mu}_0$.}
\label{fig:closed_path}
\end{figure}

The effective Lagrangian in this way can be given as
\begin{eqnarray}
\mathcal{L}_{\mathrm{eff}}(x)&=&\frac{i}{2} \int_0^{\infty}~ \frac{ds}{s}~e^{-i m^2s}~ \langle C(x_0); s \rangle^{\prime} \times \left\langle 2 \left(\cos \bar{\bm{X}}'_{+}(s) +\cos \bar{\bm{X}}'_{-}(s) \right) \right\rangle'.~~~ \label{effective Lagrangian 2}
\end{eqnarray}
For a closed path $C(x_0)$, we expand $x^{\mu}(s')$ as 
\begin{eqnarray}
C(x_0):~x^{\mu}(s')=x_0^{\mu} + \frac{1}{\sqrt{s}} \sum_{n=1}^{\infty} \left(a^{\mu}_n e^{-2\pi n i (s'/s)}+a^{\mu \dagger}_n e^{2\pi ni(s'/s)} \right),
\end{eqnarray}
where $a^{\mu}_n$ is a complex number.  

Next, we consider how to choose the origin of the harmonic oscillator potential.  Looking at the Lagrangian $A$ in Eq.~(\ref{A(s')}), the harmonic oscillator potential has an extremum (maximum or minimum) at $x^{\mu}=0$.  However, the contribution of the closed path $C(x_0)$ to the effective action should be the same for any choice of the center of the mass coordinate $x_0^{\mu}$.  To resolve this point, the ambiguity can be used that already exists in rewriting the vector potential in terms of the field strength; namely,
\begin{eqnarray}
A_{\mu}(x)=\frac{1}{2} (x^{\nu}-c^{\nu}) F_{\nu\mu}
\end{eqnarray}
is valid with any constant $c^{\mu}$.  


Now, we can understand that in order to guarantee homogeneity in space, the Hamiltonian and the Lagrangian should be modified, by replacing 
\begin{eqnarray}
x^{\mu} \to x^{\mu}-x^{\mu}_0,
\end{eqnarray}
when we consider the contribution of $C(x_0)$ to the effective action.

Then, the action variable $\bar{A}(s)$ of the point particle for the closed path $C(x_0)$ becomes $x^{\mu}_0$ independent, giving
\begin{eqnarray}
\bar{A}_{C(x_0)}(s)\equiv \int_{0~C(x_0)}^s~ ds'~A(s')=2\sum_{n=1}^{\infty} a^{\mu\dagger}_n {\ell_{\mu}}^{\nu}(n) ~a^{\nu}_n,
\end{eqnarray}
where 
\begin{eqnarray}
{\ell_{\mu}}^{\nu}(n)=-\left(\frac{\pi n}{s}\right)^2 {g_{\mu}}^{\nu}- i\left(\frac{\pi n}{s}\right) {(g_V \bm{F})_{\mu}}^{\nu} - \frac{1}{2} {{(g_A\bm{F})^2}_{\mu}}^{\nu},
\end{eqnarray}
where the matrix notation for the field strength is introduced by ${(\bm{F})_{\mu}}^{\nu}={F_{\mu}}^{\nu}$. The matrix has the Lorentz indices, the down index $\mu$ as a row index and the upper index $\nu$ as a column index.  Diagonalization of the matrix $\bm{F}$ is well known in deriving the H-E formula in QED (See Appendix A for the more details).

The four eigenvalues of $\bm{F}$ are demoted by $\{F'_{(\lambda)}\}(\lambda=\pm 1, \pm 2)$:
\begin{eqnarray}
&&F'_{(\pm 1)}=\pm F^{(1)},~\mathrm{and}~F'_{(\pm 2)}=\pm F^{(2)}, \label{eigen-values} \mathrm{with} \\
&&F^{(1)}=i Re X=\frac{i}{\sqrt{2}} \left[ (\mathcal{F}+i \mathcal{G})^{1/2} + (\mathcal{F}-i \mathcal{G})^{1/2} \right],  \label{eigen-values1}\\
&&F^{(2)}=- Im X=\frac{i}{\sqrt{2}} \left[ (\mathcal{F}+i \mathcal{G})^{1/2} - (\mathcal{F}-i \mathcal{G})^{1/2} \right],  \label{eigen-values2}
\end{eqnarray}
where
\begin{eqnarray}
X&=&\sqrt{2(\mathcal{F}+i \mathcal{G})},
\end{eqnarray}
and $\mathcal{F}$ and $\mathcal{G}$ are given in Eq.~(\ref{F and G}). 
Corresponding to the diagonalization of $\bm{F}$, the matrix $\bm{\ell}(n)$ is also diagonalized as $\{\ell (n, \lambda)\}(\lambda=\pm1, \pm2)$:
\begin{eqnarray}
\ell (n, \lambda)&=&-\left(\frac{\pi n}{s}\right)^2-i \left(\frac{\pi n}{s}\right) (g_V F'_{(\lambda)}) - \frac{1}{2} (g_A F'_{(\lambda)})^2 \nonumber \\
&=&-\left(\frac{\pi n}{s}+ig_{+} F'_{(\lambda)}\right) \left(\frac{\pi n}{s}+ig_{-} F'_{(\lambda)}\right),
\end{eqnarray}
where $g_{\pm}$ are given by
\begin{eqnarray}
g_{\pm}=\frac{1}{2} (g_V \pm \sqrt{g_V^2+2g_A^2}). \label{g_pm}
\end{eqnarray}

Now, the transition amplitude (\ref{transition amplitude 2}) can be estimated by the Gaussian integrations as
\begin{eqnarray}
&&\langle C(x_0); s \rangle \equiv \int_{C(x_0)} \mathcal{D} x^{\mu}(s')~ e^{i\int_{0~C(x_0)}^s ds' L \left(x(s'), ~\dot{x}(s')\right)}  \\
&&=\prod_{n=1}^{\infty}\prod_{\mu=0-3}\int \mathcal{D} a^{\mu \dagger}_{n} ~\mathcal{D} a_{\mu n} ~e^{i \left(2\sum_{n=1}^{\infty} a^{\mu\dagger}_n~ {\ell_{\mu}}^{\nu}(n) ~a_{\nu n} \right) } \propto \prod_{\lambda=\pm1, \pm2} \prod_{n=1}^{\infty} \frac{1}{\ell(n, \lambda)}.~~~~
\end{eqnarray}

So far we haven't cared about the numerical factors, since they are common with or without the external fields $\bm{F}$ or its eigenvalues $F'$. 
For this point we will again follow Schwinger \cite{Schwinger}. Then,
\begin{eqnarray}
\langle C(x_0); s \rangle_{F' \ne 0} = \langle C(x_0); s \rangle_{F'=0} \times \frac{\langle C(x_0); s \rangle_{F'\ne 0}}{\langle C(x_0); s \rangle_{F'=0}}, \label{product}
\end{eqnarray}
and we estimate $\langle C(x_0); s \rangle_{F'=0}$ more explicitly, and require the second factor to become 1 when the external fields disappear. 

To obtain the first factor, we use the Schr\"odinger equation for the transition amplitude from $x''$ to $x'$,
\begin{eqnarray}
\left( i \frac{\partial}{\partial s} -\hat{H}_{x'} \right) \langle x'(s)| x''(0) \rangle_{F'=0}=0,
\end{eqnarray}
where $\hat{H}_{x'}=\partial'_{\mu} \partial'^{\mu}$, and $\langle x'(s) | x''(0) \rangle \to \delta^{(4)}(x'-x'')$ should be satisfied when $s \to 0$.  This can be easily solved and we have
\begin{eqnarray}
\langle x'(s)| x''(0) \rangle_{F'=0}=\frac{-i}{(4\pi s)^2} e^{-\frac{i}{4s}(x'-x'')^2},  \mathrm{and~hence}~\langle x(s)| x(0) \rangle_{F'=0}=\frac{-i}{(4\pi s)^2}.
\end{eqnarray}
The result does not depend on $x^{\mu}$, which implies that $\langle C(x_0); s \rangle=\frac{-i}{(4\pi s)^2}$.

As for the second factor in Eq.~(\ref{product}) it yields \footnote{$\frac{\sinh x}{x}= \prod_{n=1}^{\infty}~\left(1+ \left(\frac{x}{\pi n}\right)^2\right)$.}
\begin{eqnarray}
&&\frac{\langle C(x_0); s\rangle_{F'\ne 0}}{\langle C(x_0); s \rangle_{F'=0}} =\prod_{n=1}^{\infty} ~\prod_{\lambda=\{\pm1, \pm2\}} \left(\frac{\frac{\pi n}{s}}{\frac{\pi n}{s}+i g_{+}F'_{(\lambda)}}\right) \left( \frac{\frac{\pi n}{s}}{\frac{\pi n}{s}+i g_{-}F'_{(\lambda)}} \right) \\
&=&\frac{F^{(1)}F^{(2)}(g_{+}s)^2}{\sinh(g_{+}F^{(1)}s) \sinh(g_{+}F^{(2)}s)}
\times \frac{F^{(1)}F^{(2)}(g_{-}s)^2}{\sinh(g_{-}F^{(1)}s) \sinh(g_{-}F^{(2)}s)}  \label{the ratio 1}
\\ 
&=&\frac{-2i (g_{+}s)^2\mathcal{G}}{\cosh(g_{+}X (is))- \cosh(g_{+}X^{\dagger} (is))}
\times \frac{-2i (g_{-}s)^2\mathcal{G} }{\cosh(g_{-}X (is))-\cosh(g_{-}X^{\dagger} (is))},~~~~\label{the ratio 2}
\end{eqnarray}
where $\mathcal{G}=\mathrm{Re}X \mathrm{Im}X$.

Now, by performing the Wick rotation $s \to -is$ and using Eq.~(\ref{the ratio 1}), we obtain the following effective Lagrangian:
\begin{eqnarray}
\mathcal{L}_{\mathrm{eff}}&=&-\frac{1}{8\pi^2} \int_0^{\infty} \frac{ds}{s^3} e^{-m^2 s} \frac{(g_{+}s)^2 \mathcal{G}}{\mathrm{Im} \cosh(g_{+}Xs)} \times \frac{(g_{-}s)^2 \mathcal{G}}{\mathrm{Im} \cosh(g_{-}Xs)} \nonumber \\
&&~~~~~~~~~~~~~\times \frac{1}{2} \left\langle \left(\cos \bar{\bm{X}}'_{+}(s \to -is) +\cos \bar{\bm{X}}'_{-} (s \to -is)\right) \right\rangle'. \label{generalized H-E}
\end{eqnarray}
This is a main result of this paper, and the formula may be called the ``generalized H-E formula,'' since it is possible to include the parity-violating interactions. 

The formula is quite similar to that in QED, but has the complexity in which the expectation value should be taken for the last factor relevant to the spin averaging.  More precisely let us compare the formula Eq.~(\ref{generalized H-E}) to the H-E formula in QED; for example, to Eq.~(3.44) in Ref.~\cite{Schwinger} .  For this purpose, we choose $g_V=-e$ and $g_A = 0$.  Then, the  spin-dependent last factor, which is evaluated by the expectation value in the general case of Eq.~(\ref{generalized H-E}), becomes a simple factor $\frac{1}{2} (\cos (-is) (-e) X + \cos (-is)(-e) X^{\dagger})=\mathrm{Re} (\cosh (esX))$, which is identical to the numerator of Eq.~(3.44) in Ref.~\cite{Schwinger}.  In Eq.~(\ref{generalized H-E}), there exist two factors depending on $g_{+}$ and $g_{-}$, which shows that there exist two sets of harmonic oscillators, $(x, p_x)$ and $(y, p_y)$, in the general case with $g_A\ne0$.  In QED one of the factors disappears because of $g_{+}=0$, which shows that only one set of harmonic oscillators, $(p_x, p_y)$, remains as the cyclotron motion in QED.

If the parity-violating interaction exists with the axial vector coupling $g_A$, however, the spin-dependent factor and the spin-independent factor interact with each other.  This interaction refuses a simple expression, and the formula is expressed in terms of the perturbation theory with the field -dependent propagator $\Delta(s'-s'')_{\alpha\beta}$ which will be given in the next section. Therefore, the generalized formula is expressed in terms of the expectation value $\langle \cdots \rangle'$ averaged with the field-dependent propagators.  

\section{Perturbation theory and the effective Lagrangian, 
$a\mathcal{F}^2+b\mathcal{G}^2+ ic\mathcal{F}\mathcal{G}$ }  
If we estimate Eq.~(\ref{generalized H-E}) at the second order in $\mathcal{F}$ and $\mathcal{G}$,  we can obtain the generalized effective Lagrangian $\mathcal{L}^{(2)}_{\mathrm{eff}}=a\mathcal{F}^2+b\mathcal{G}^2+ ic\mathcal{F}\mathcal{G}$ which is the fourth order in the background fields $\bm{E}$ and $\bm{B}$, and is $O((g_{V}/m\; \mathrm{or}\;g_{A}/m)^4)$. 

In our case with parity violation, perturbative calculation is necessary, so we will first summarize the Feynman rules, (P) propagator and (V) vertices.

The propagator is originally defined in matrix form in Eq.~(\ref{def of the propagator}), but after diagonalizing the background field $\bf{F}_{\mu}^{~\nu}$, the propagator is also diagonalized, and,  corresponding to the eigenvalue $F'_{(\lambda)}$ of $\bf{F}_{\mu}^{~\nu}$, the propagator takes its eigenvalue $\Delta(s'-s'')_{(\lambda)}$, which reads
\begin{eqnarray}
&&\Delta(s'-s'')_{(\lambda)} \nonumber \\
&&=\frac{-1}{4(g_{+}-g_{-})F'_{(\lambda)}}
\left\{ \frac{e^{-2g_{+} F'_{(\lambda)} (s'-s''-\epsilon(s'-s'')\frac{s}{2})}}{\sinh (g_{+}F'_{(\lambda)}s)}-\frac{e^{-2g_{-} F'_{(\lambda)} (s'-s''-\epsilon(s'-s'')\frac{s}{2})}}{\sinh (g_{-}F'_{(\lambda)}s)} \right\}, \label{propagator}~~~
\end{eqnarray}
where $\epsilon(s'-s'')$ is a step function, 
\begin{eqnarray}
\epsilon(x)= 1 ~\mathrm{for}~ x>0, ~ -1 ~\mathrm{for}~ x<0,~\mathrm{and}~0 ~\mathrm{for}~ x=0, 
\end{eqnarray}
and $g_{\pm}$ is defined in Eq.~(\ref{g_pm}).

The transformation from the usual Lorentz frame to the ``diagonal frame'' (the frame in which $\bm{F}_{\mu}^{~\nu}$ is diagonalized) is a little complicated, so we refer to Appendix A on this issue.  In the real calculation, it is enough to understand that keep the matrix form as far as possible, and diagonalize the $\bm{F}$ and the propagator at the last moment.  We denote the component of the diagonal frame as $(\mu), (\nu), (\alpha), (\beta), \ldots$ Equation~(\ref{propagator}) is derived so as to be periodic in $s'-s''$ with a period $s$, since the dynamics is given in the finite time interval $s$.  Therefore, for the propagator we have the following:
\begin{eqnarray}
&&(P1) ~\mathrm{(Transformation~ from~ the ~Lorentz ~frame} \to \mathrm{the~diagonal ~frame)}: \nonumber \\
&&~~~~\langle x_{\mu}(s') x^{\nu}(s'') \rangle' =2i~ \Delta(s'-s'')_{\mu}^{~\nu} \to 2i~\Delta(s'-s'')_{(\mu)}^{~(\nu)}=2i~ \delta_{(\lambda)}^{~(\nu)} \Delta(s'-s'')_{(\mu)}^{~(\lambda)}, \nonumber \\
&&~~~~\langle x_{\mu}(s') x_{\nu}(s'') \rangle' = 2i~ \Delta(s'-s'')_{\mu\nu} \to 2i~\Delta(s'-s'')_{(\mu)(\nu)} =2i~ \Delta(s'-s'')_{(\mu)}^{~(\lambda)} g_{(\lambda)(\nu)};~~~~~~ \\
&&(P2): ~(\mathrm{Periodicity}): \nonumber \\
&&~~~~\Delta(s'-s'')_{(\mu)}^{~(\nu)}=\Delta(s'-s''-ns)_{(\mu)}^{~(\nu)}, \Delta(s'-s'')_{(\mu)(\nu)}=\Delta(s'-s''-ns)_{(\mu)(\nu)}, \nonumber \\
&&~~~~\mathrm{since}~ \Delta(s', s'')_{(\lambda)}=\Delta(s'-s''-ns)_{(\lambda)}  (n=\mathrm{any~integer});\\
&&(P3): ~(\mathrm{Transposition}): \nonumber \\
&&~~~~\Delta(s'-s'')^{\mu}_{~\nu}=\Delta(s''-s')_{\nu}^{~\mu}, \Delta(s'-s'')_{\mu\nu}=\Delta(s''-s')_{\nu\mu}, 
\end{eqnarray}
where in $(P3)$, the minus sign arising from the antisymmetric nature of $\bf{F}$ is compensated by the change of the time direction.

What always appears in the following calculation is $\frac{\partial}{\partial s'}\Delta(s', s'')_{(\lambda)}$, so it is convenient to introduce the notation 
$\Delta(\dot{s'}, s'')_{(\lambda)} \equiv \frac{\partial}{\partial s'}\Delta(s', s'')_{(\lambda)}=\frac{\partial}{\partial s'}\Delta(s'-s'')_{(\lambda)}$.   Explicitly, we have
\begin{eqnarray}
&&(P4)~(\mathrm{The ~first~derivative ~of ~the ~propagator}): \nonumber \\
&&~~~\Delta(\dot{s'}, s'')_{(\lambda)} \equiv \frac{\partial}{\partial s'}\Delta(s', s'')_{(\lambda)} \nonumber \\
&&~~~~=\frac{1}{2(g_{+}-g_{-})} \times \left\{ \frac{g_{+} \left(e^{-2g_{+} F'_{(\lambda)} (s'-s''-\epsilon(s'-s'')\frac{s}{2})}-s \delta(s'-s'')\right)}{\sinh (g_{+}F'_{(\lambda)}s)} \right. \nonumber \\
&& \left. ~~~~~~~~~~~~~~~~~~~~~-\frac{g_{-} \left(e^{-2g_{-} F'_{(\lambda)} (s'-s''-\epsilon(s'-s'')\frac{s}{2})}-s \delta(s'-s'')\right)}{\sinh (g_{-}F'_{(\lambda)}s)} \right\}. \label{derivative of propagator}
\end{eqnarray}

The transposition of the dotted propagator is
\begin{eqnarray}
&&(P5): ~(\mathrm{Transposition~of~the~dotted~propagator}): \nonumber \\
&&~~~~\Delta(\dot{s'}, s'')^{\mu}_{~\nu}=-\Delta(\dot{s''}, s')_{\nu}^{~\mu}, \Delta(\dot{s'}, s'')_{\mu\nu}=-\Delta(\dot{s''}-s')_{\nu\mu},\nonumber \\
&&~~~~\Delta(\dot{s'}, s'')_{(\lambda)} \nonumber=-\Delta(\dot{s''}, s')_{(-\lambda)},
\end{eqnarray}
where the transposition of the antisymmetric matrix $\bm{F}$ gives $(\lambda)\to (-\lambda)$.

The proper time $s'$, $s'', \ldots$ are always integrated from 0 to $s$, and the propagator has the periodicity $(P2)$, so that, by using integration by parts, we can perform the following change:
\begin{eqnarray}
&&(P6)~\mathrm{(Change ~of ~the ~position ~of~differentials)}~~ \mathrm{For ~example,}~\nonumber \\
&&~~~~\int_0^s ds'ds'' \cdots \{\cdots \Delta(\dot{s'}, \dot{s''}) \cdots \Delta(s'', s''') \cdots\} \nonumber \\
&&~~~~\to \int_0^s ds'ds'' \cdots -\{\cdots \Delta(\dot{s'}, s'') \cdots \Delta(\dot{s''}, s''') \cdots\}.
\end{eqnarray}
With the help of this $(P6)$, all the propagators in our problem are changed to  $\Delta(\dot{s'}, s'')$. 

Next, we examine the vertices.  The vertex function can be extracted from Eq.~(\ref{B(s')}).  $B_{\mu\nu}(s')$ consists of the $x(s')$-dependent term and the constant term:
\begin{eqnarray}
B_{\mu\nu}(s')&=&{V^{(1)}_{\mu\nu}}^{~\alpha\beta} \dot{x}_{\alpha}(s') x_{\beta}(s')+V^{(0)}_{\mu\nu}, \\
&&{V^{(1)}_{\mu\nu}}^{~\alpha\beta}=\frac{1}{2}g_A \epsilon_{\mu\nu}^{~~~\alpha\beta'}F_{\beta'}^{~\beta}, \label{V(1)} \\
&&V^{(0)}_{\mu\nu}=-(g_V F_{\mu\nu}-ig_A \tilde{F}_{\mu\nu}). \label{V(0)}
\end{eqnarray}
Therefore, the vertex functions can be denoted as
\begin{eqnarray}
&&(V1) ~(\mathrm{vertex ~functions}) \nonumber \\
&&~~~~ (V^{(1)}):~ V_{\mu\nu,}^{(1)~~~\dot{\alpha}\beta},~\mathrm{and} ~(V^{(0)}):~V^{(0)}{\mu\nu}.
\end{eqnarray}
where the ``dot'' may be added on the index $\alpha$ if we wish to clarify that differentiation by $s'$ is associated with the external line $x_{\alpha}(s')$.  The dot may also be depicted on the propagator to show the derivative is on this line, and the arrow from $s'' \to s'$ is written on the propagator $2i \Delta(\dot{s'}, s'')_{\alpha\beta}$; the dotted external line shows the flow of the index $(\mu\nu)$ from outside.  We have to remember that all the vertices are integrated over the proper time. These vertices are depicted in Fig.~2.

\begin{figure}[ht]
\center
\includegraphics[width=12cm]{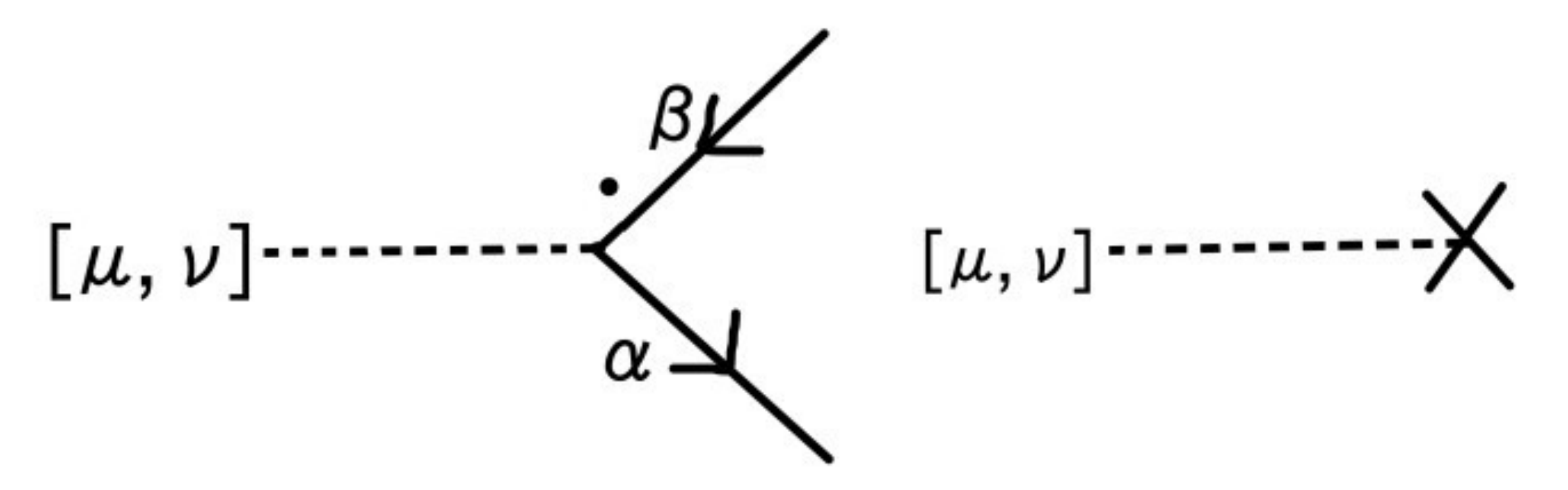}
\caption{Vertex functions}
\label{fig:vertices}
\end{figure}

What we need to estimate, if we want to know the first nontrivial effective Lagrangian, are the second-order terms in $\mathcal{F}$ and $\mathcal{G}$, or the fourth-order terms in $\bar{\bm{X}}'_{\pm}(s)$ should be estimated for the following expectation value: 
\begin{eqnarray}
&&\frac{1}{2} \left\langle \left(\cos \bar{\bm{X}}'_{+}(s) +\cos \bar{\bm{X}}'_{-} (s)\right) \right\rangle' \nonumber \\
&&=1-\frac{1}{4} \left\langle (\bar{\bm{X}}'_{+} (s))^2 + (\bar{\bm{X}}'_{-} (s))^2 \right\rangle'+ \frac{1}{48} \left\langle (\bar{\bm{X}}'_{+} (s))^4 +  (\bar{\bm{X}}'_{-} (s))^4 \right\rangle' + \cdots \\
&&=1-\langle \bar{\bm{\mathcal{F'}}}(s) \rangle' + \frac{1}{6} \langle (\bar{\bm{\mathcal{F'}}}(s))^2-(\bar{\bm{\mathcal{G'}}}(s))^2 \rangle' + \cdots. \label{Hosei}
\end{eqnarray}
Here, we remember that $\bar{\bm{\mathcal{F'}}}(s)= \frac{1}{4} \bar{B}_{\mu\nu}(s)\bar{B}^{\mu\nu}(s), ~~\bar{\bm{\mathcal{G'}}}(s)= \frac{1}{4} \bar{B}_{\mu\nu}(s)\bar{\tilde{B}}^{\mu\nu}(s)$ in Eq.~(\ref{F and G by B(s')}).

First we show that the tadpole diagrams vanish as in Fig.~3, since they become the total derivative:
\begin{eqnarray}
\int_0^s  ds' ~{V^{(1)}_{\mu\nu,}}^{~\alpha\beta} ~2i\frac{\partial}{\partial s'} \Delta(s')_{\alpha\beta}=0.
\end{eqnarray}

\begin{figure}[ht]
\center
\includegraphics[width=8cm]{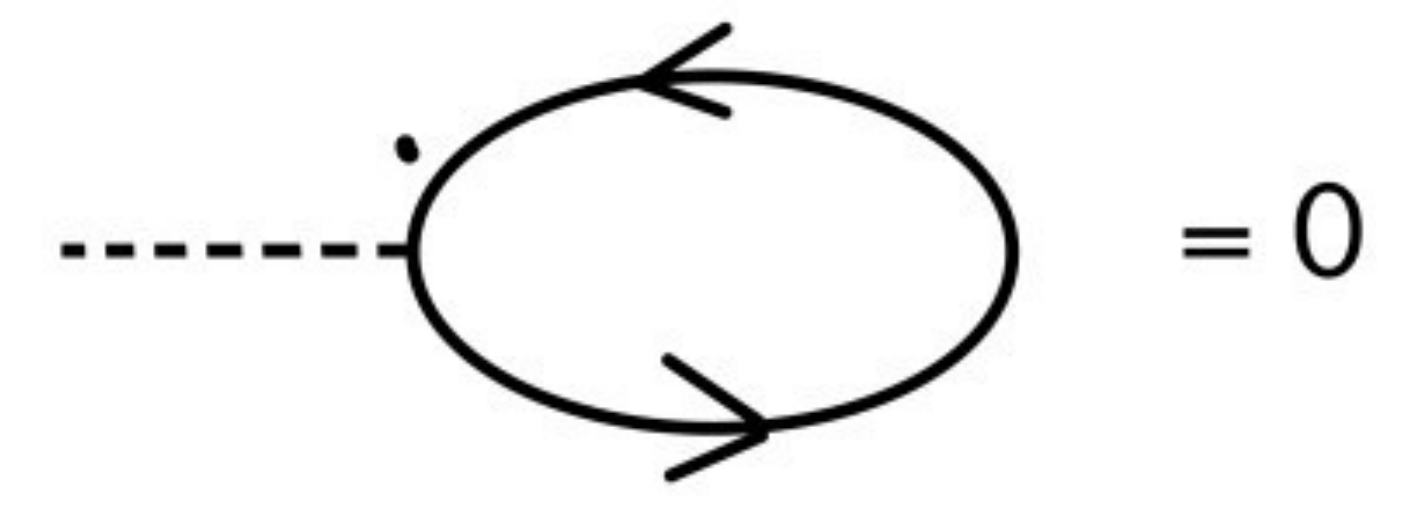}
\caption{Tadpole diagrams}
\label{fig:tadpole}
\end{figure}

Without the tadpole diagrams, the relevant Feynman diagrams at the fourth order, having four vertices, are depicted in Fig.~4). 

\begin{figure}[ht]
\center
\includegraphics[width=14cm]{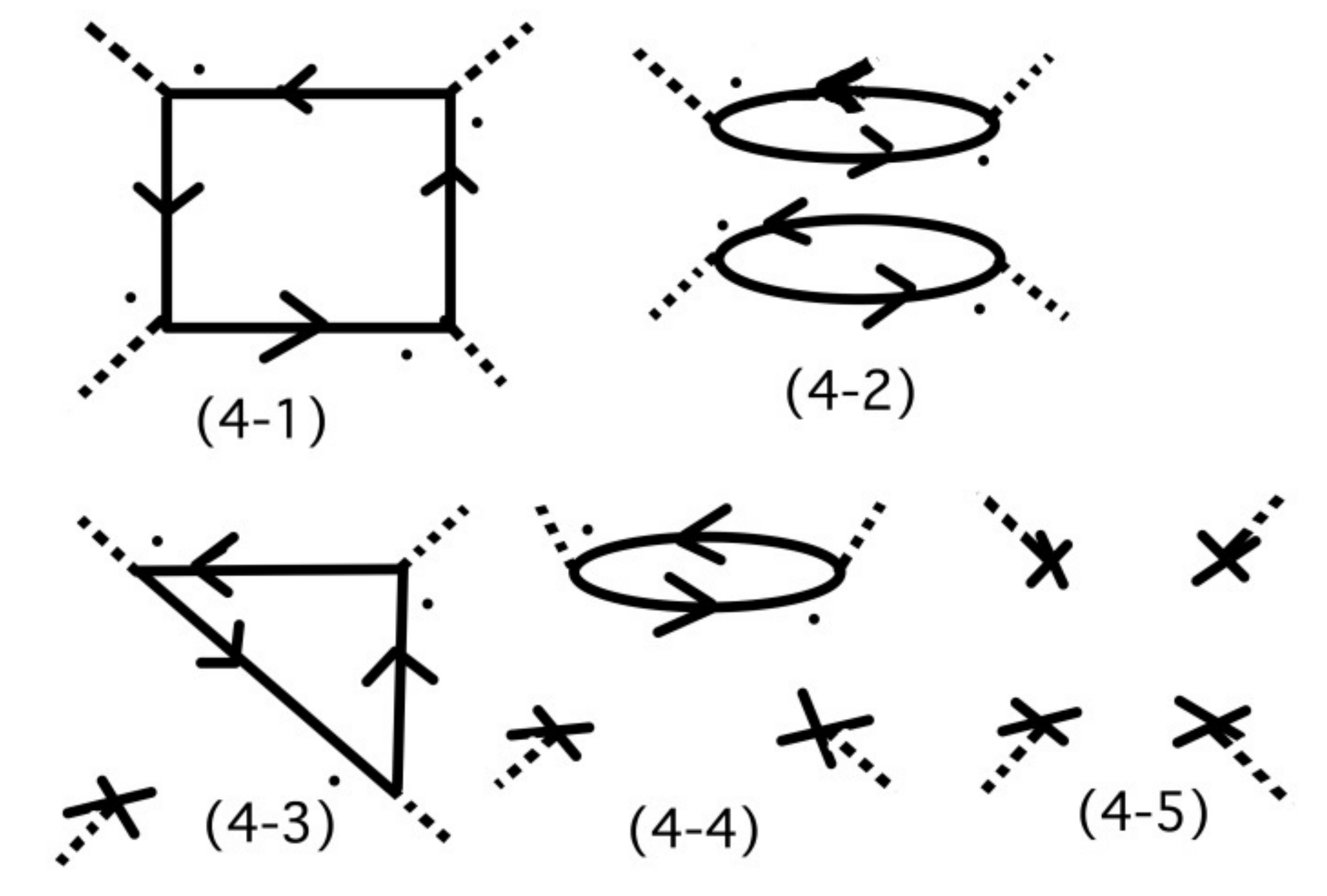}
\caption{Feynman diagrams at the fourth order}
\label{fig:4thorder}
\end{figure}

Here, the propagators are so arranged as to have one time derivative, by using $(P5)$.  If there exist two derivatives on a propagator, then one of the derivatives can be moved to the other external line, which is not involved in the relevant propagator.  After reducing all the propagators to $\Delta(\dot{s}, s'')$, the vertex function is modified to the ``antisymmetrized'' effective vertex,

\begin{eqnarray}
(V2)~\mathrm{(Effective ~vertex ~for ~the~first~derivative~propagators)} \nonumber \\
-{V^{(1)}_{\mu\nu}}^{~[\alpha, \beta]} \equiv - (V_{\mu\nu}^{~~~\alpha\dot{\beta}} -V_{\mu\nu}^{~~~\dot{\beta}\alpha}), \label{effective V(1)}
\end{eqnarray}
since the two different contractions at a vertex can be combined to give the following result, via integration by parts:
\begin{eqnarray}
&&2i \Delta (\cdot, s')_{\cdot\beta}~  V_{\mu\nu}^{~~~\dot{\alpha}\beta}~2i \Delta (\dot{s'}, \cdot)_{\dot{\alpha} \cdot} + 2i \Delta (\cdot, \dot{s'})_{\cdot \dot{\alpha}} ~V_{\mu\nu}^{~~~\dot{\alpha}\beta}~2i \Delta (s', \cdot)_{\beta\cdot}  \\
&&=2i \Delta (\cdot, s')_{\cdot\beta}~  V_{\mu\nu}^{~~~\dot{\alpha}\beta}~2i \Delta (\dot{s'}, \cdot)_{\dot{\alpha} \cdot} - 2i \Delta (\cdot, s')_{\cdot \dot{\alpha}} ~ V_{\mu\nu}^{~~~\dot{\alpha}\beta}~2i \Delta (\dot{s'}, \cdot)_{\beta \cdot} \nonumber \\
&&= 2i \Delta (\cdot, s')_{\cdot \alpha} ~(-{V^{(1)}_{\mu\nu}}^{~[\alpha, \beta]})~ 2i \Delta (\dot{s'}, \cdot)_{\beta \cdot}
\end{eqnarray}

Then, the loop diagram $(LP)$ with $N$ external legs having indices $(\mu_1\nu_1), \ldots,  (\mu_N\nu_N)$ yields
\begin{eqnarray}
&&LP_{(\mu_1\nu_1, \mu_2\nu_2, \cdots, \,u_N\nu_N)} =(-2i)^N \int_0^s ds_1 ds_2 \cdots ds_N \nonumber \\
&&\times \left\{ V_{\mu_1\nu_1}^{(1)[\alpha_1\beta_1]} \Delta(\dot{s_1}, s_2)_{\beta_1\alpha_2} V_{\mu_2\nu_2}^{(1)[\alpha_2\beta_2]} \Delta(\dot{s_2}, s_3)_{\beta_2\alpha_3} \cdots V_{\mu_N\nu_N}^{(1)[\alpha_N\beta_N]}\Delta(\dot{s_N}, s_1)_{\beta_N\alpha_1} \right\}~~~~~~ \\
&&=(-2i)^N \int_0^s ds_1 ds_2 \cdots ds_N \nonumber \\
&&\times \left\{ {V^{(1)}_{\mu_1\nu_1}}_{[\alpha_1}^{~~\beta_1]} \Delta(\dot{s_1}, s_2)_{\beta_1}^{~~\alpha_2} {V^{(1)}_{\mu_2\nu_2}}_{[\alpha_2}^{~~\beta_2]} \Delta(\dot{s_2}, s_3)_{\beta_2}^{~~\alpha_3} \cdots {V^{(1)}_{\mu_N\nu_N}}_{[\alpha_N}^{~~\beta_N]}\Delta(\dot{s_N}, s_1)_{\beta_N}^{~~\alpha_1} \right\} \\
&&=(-2i)^N  \left\{ {V^{(1)}_{\mu_1\nu_1}}_{[(\alpha_1)}^{~~(\alpha_2)]} {V^{(1)}_{\mu_2\nu_2}}_{[(\alpha_2)}^{~~(\alpha_3)]} \cdots {V^{(1)}_{\mu_N\nu_N}}_{[(\alpha_N)}^{~~(\alpha_1)]}  \right\} \nonumber \\
&&\times \int_0^s ds_1 ds_2 \cdots ds_N \left\{ \Delta(\dot{s_1}, s_2)_{(\alpha_2)} \Delta(\dot{s_2}, s_3)_{(\alpha_3)} \cdots \Delta(\dot{s_N}, s_1)_{(\alpha_1)} \right\}, \label{LP}
\end{eqnarray}
by taking the ``diagonal frame'' in the last equation.

The estimation of $\langle \bar{\bm{\mathcal{F'}}}(s) \rangle' $, $\langle (\bar{\bm{\mathcal{F'}}}(s))^2\rangle'$, and $\langle (\bar{\bm{\mathcal{G'}}}(s))^2 \rangle' $ using the propagator and the vertices is summarized in Appendix~C, and gives
\begin{eqnarray}
\langle \bar{\bm{\mathcal{F'}}}(s) \rangle'&=& \frac{1}{3}~g_A^2\left\{
2 s^2 \mathcal{F} + \frac{1}{15} s^4 \left( (4g_V^2+ 2g_A^2 ) \mathcal{F}^2 - 3 g_V^2 \mathcal{G}^2 \right) \right\} \nonumber \\
&&+ s^2 \left((g_V^2+g_A^2)\mathcal{F} - 2i g_Vg_A \mathcal{G}\right), \label{mathcal F}
\end{eqnarray}

\begin{eqnarray}
\langle (\bar{\bm{\mathcal{F'}}}(s))^2\rangle'=&&\frac{1}{16} s^4 \left\{ \frac{32}{9}g_A^4 (3\mathcal{F}^2-\mathcal{G}^2) + \frac{8}{45} g_A^4  (5\mathcal{F}^2+\mathcal{G}^2) \right. \nonumber \\
&&~~~~~~+\frac{64}{3}g_A^2 \left( (g_V^2+g_A^2)\mathcal{F}^2 -2i g_Vg_A \mathcal{F}\mathcal{G} \right) \nonumber \\
&&~~~~~~\left. +16 \left( (g_V^2+g_A^2)\mathcal{F} -2i g_Vg_A\mathcal{G} \right)^2\right\}, \label{mathcal F squared} 
\end{eqnarray}
and
\begin{eqnarray}
\langle (\bar{\bm{\mathcal{G'}}}(s))^2\rangle'=&&\frac{1}{16} s^4 \left\{ \frac{32}{9}g_A^4 (-\mathcal{F}^2+3\mathcal{G}^2) + \frac{8}{45} g_A^4  (\mathcal{F}^2+5\mathcal{G}^2) \right. \nonumber \\
&&~~~~~~+\frac{64}{3}g_A^2 \left( (g_V^2+g_A^2)\mathcal{G}^2 + 2i g_Vg_A \mathcal{F}\mathcal{G} \right) \nonumber \\
&&~~~~~~\left. +16 \left( (g_V^2+g_A^2)\mathcal{G} + 2i g_Vg_A\mathcal{F} \right)^2\right\},  \label{mathcal G squared}
\end{eqnarray}
where a symmetry (duality symmetry between electricity and magnetism) $\mathcal{F} \to \mathcal{G}$ and $\mathcal{G} \to -\mathcal{F}$ appears between $\langle (\bar{\bm{\mathcal{F'}}}(s))^2\rangle'$ and $\langle (\bar{\bm{\mathcal{G'}}}(s))^2\rangle'$, since it exists between $B_{\mu\nu}(s')$ and $\tilde{B}_{\mu\nu}(s')$.

Now the factor in Eq.~(\ref{generalized H-E}) expressed in terms of the expectation values is obtained by Eq.~(\ref{Hosei}) and Eqs.(\ref{mathcal F})-(\ref{mathcal G squared}).  By performing the expansion in $s$ of the integrand of the effective Lagrangian Eq.~(\ref{generalized H-E}) except for the factor $s^{-3}e^{-m^2s}$, and carrying out the integration over $s$, we obtain the effective Lagrangian as $\mathcal{L}_{\mathrm{eff}}=\mathcal{L}^{(0)}_{\mathrm{eff}}+\mathcal{L}^{(1)}_{\mathrm{eff}}+\mathcal{L}^{(2)}_{\mathrm{eff}} + \cdots $, where the suffix $(n)$ indicates that the corresponding integrand behaves as $s^{2n}$.  Then, the $O(s^4)$ terms in the integrand (except for the factor $s^{-3}e^{-m^2s}$) give the effective Lagrangian of the fourth order $O(\bm{F}^4)$ in the external fields.

Therefore, the generalized H-E formula is derived as $\mathcal{L}_{\mathrm{eff}}=a\mathcal{F}^2 + b \mathcal{G}^2 + ic \mathcal{F}\mathcal{G}$ with the following coefficients:
\begin{eqnarray}
&&a=\frac{1}{(4\pi)^2m^4}\left(\frac{8}{45}~ g_V^4- \frac{4}{5}~g_V^2g_A^2 - \frac{1}{45}~g_A^4\right), \label{a} \\
&&b=\frac{1}{(4\pi)^2m^4}\left (\frac{14}{45} ~g_V^4 + \frac{34}{15}~ g_V^2 g_A^2 + \frac{97}{90}~g_A^4 \right), \label{b} \\
&&c=\frac{1}{(4\pi)^2 m^4} \left(\frac{4}{3}~g_V^3g_A + \frac{28}{9} ~g_V g_A^3 \right), \label{c}
\end{eqnarray}

If we set $g_V=-e$ and $g_A=0$, the original H-E formula Eq.~(\ref{H-E}) is reproduced. 

In case of $V+A$ or $V-A$ interactions, we set $g_A= \pm g_V$, respectively, and we have
\begin{eqnarray}
&&a(V \pm A)=-\frac{29}{45}\left(\frac{g_V^2}{4\pi m^2}\right)^2,~ b(V \pm A)=\frac{329}{90}\left(\frac{g_V^2}{4\pi m^2}\right)^2, ~\mathrm{and}\nonumber \\
&&c(V \pm A)=\pm \frac{40}{9}\left(\frac{g_V^2}{4\pi m^2}\right)^2. \label{V+-A}
\end{eqnarray}

Here, $\mathcal{F}^2$ and $\mathcal{G}^2$ do not violate the parity invariance, while $\mathcal{F}\mathcal{G}$ does violate the parity invariance, so that the coefficients $a$ and $b$ describe the usual parity-conserving effects, while $c$ gives the parity-violating effects. It is also noted that the contribution of $\mathcal{F}\mathcal{G}$ in the effective action is purely imaginary.  


\section{Conclusion and discussion}


In this paper we have derived the generalized Heisenberg-Euler formula Eq.~(\ref{generalized H-E}), which is applicable for parity-violating theories.  
So, it may afford a basis for experiments with a laser beam to probe the parity-violating effects in the dark sector, if the photon mixes with the gauge field in the dark sector. See Ref.~\cite{application to OVAL} for the physical interpretation of the formula and the possible experimental methods.

One thing that is foreign from the usual H-E formula in QED is that one factor in the formula is given in terms of the expectation value denoted by $\langle \cdots \rangle'$ as in Eq.~(\ref{generalized H-E}), since the spin-independent part and the spin-dependent part are mixed in the presence of the axial gauge coupling.  This may be understood from the following form of the gauge interaction with the vector ($g_V$) as well as the axial vector ($g_A $) couplings,
\begin{eqnarray}
L_{\mathrm{gauge~ interaction}}=\dot{x}^{\mu} \left\{ g_V A_{\mu}(x) +i g_A \gamma_5 \sigma^{\mu\nu} A_{\nu}(x)\right\},
\end{eqnarray}
where the gauge field couples to the spin current $J^{\mu}_{\mathrm{spin}}(x)$ through the axial vector coupling, from which we can understand that the spin at time $s'$ is dynamically connected to the location $x^{\mu}(s')$ of the spin:   
\begin{eqnarray}
J^{\mu}_{\mathrm{spin}}(x)= g_A (i \gamma_5 \sigma^{\lambda\mu} )\dot{x}_{\lambda}
\end{eqnarray}
  
Therefore, it is also convenient to introduce the Grassmannian dynamical variable $\psi^{\mu}(s')$ for the spin in addition to the coordinate $x^{\mu}(s')$, and to derive the H-E formula following an analogous method to the Neveu-Schwarz-Ramond superstring theory~\cite{Ramond1,Ramond2}.
For this purpose the Lagrangian may be chosen as 
\begin{eqnarray}
&&L(x^{\mu}(s), \dot{x}^{\mu}(s), \psi^{\mu}(s), \dot{\psi}^{\mu}(s)) = -\frac{1}{4}(\dot{x}^{\mu})^2 + i \frac{1}{2} \psi_{\mu} \dot{\psi}^{\mu} \nonumber \\
&&+ \frac{1}{2} x^{\mu} \left\{g_V F_{\mu\nu} +g_A\left(i\frac{1}{2} [\psi^{\alpha}, ~ \psi^{\beta}] \tilde{F}_{\alpha\beta} +i [\psi_{\mu}, ~ \psi^{\lambda}] \tilde{F}_{\lambda\nu} \right) \right\} \dot{x}^{\nu} \nonumber \\
&&-\frac{1}{2} x^{\mu}  (g_A)^2 F_{\mu\lambda} F^{\lambda\nu} x_{\nu} - i\frac{1}{2}[\psi^{\mu},~\psi^{\nu}] \left( g_V F_{\mu\nu} -i g_A \tilde{F}_{\mu\nu}\right).~\label{Lagrangian of point particle}
\end{eqnarray}
In the above, $\gamma_5 \sigma_{\mu\nu}=-i\frac{1}{2} \epsilon_{\mu\nu\lambda\rho}\sigma^{\lambda\rho}$ is used as well as the definition of the dual field.

In this method, $\mathrm{Tr}(e^{-i \hat{H} s})$ can be written as follows:
\begin{eqnarray}
\mathrm{Tr}(e^{-i \hat{H} s})&=&\int~d^4 x \langle x(s), \psi(s) | x(0), \psi(0) \rangle \\
&=& \int d^4x\int_{x^{\mu}(0)=x^{\mu},\psi^{\mu}(0)=\psi^{\mu}(s)}^{x^{\mu}(s)=x^{\mu}, \psi^{\mu}(s)=\psi^{\mu}(0)} \mathcal{D} x^{\mu}(s') \mathcal{D} \psi^{\mu}(s')~ \nonumber\\
&& ~~~~~~~~~~e^{i\int_0^s~ ds'~L\left(x(s'), ~\dot{x}(s'), ~\psi(s'), ~\dot{\psi}(s') \right)}.~~~ \label{transition amplitude 1}
\end{eqnarray}

So far we have not discussed on the gauge invariance of the theory.  The gauge symmetry of Eq.~(\ref{original action}) is 
\begin{eqnarray}
&&\psi(x) \to \psi'(x)= e^{i (g_V + \gamma_5 g_A) \theta (x) } \psi(x), \\
&&A_{\mu}(x) \to A'_{\mu}(x) =  A_{\mu}(x) - \partial_{\mu} \theta (x).
\end{eqnarray}
The fermion mass term manifestly breaks this gauge symmetry.  In order for the action to be invariant under the gauge symmetry, we have to introduce the ``Higgs field" $\Phi(x)= \phi(x) + i \chi(x)$, or introduce the scalar $\Phi_S(x) ~(\sigma$ meson) and the pseudoscalar $\Phi_P(x) ~(\pi$ meson), and express it as $\Phi(x)= \sigma(x) + i \pi(x)$~\cite{Nambu-Goldstone1,Nambu-Goldstone2,Nambu-Goldstone3,Nambu-Goldstone4}),
\begin{eqnarray}
\mathcal{L}_{\psi}=\bar{\psi}(x) \left[\gamma^{\mu} \left( i\partial_{\mu} -(g_V+g_A \gamma_5) A_{\mu}(x)\right) - (\phi(x)+i \gamma_5 \chi(x)) \right] \psi(x). \label{gauge invariant Lagrangian}
\end{eqnarray}
If the Higgs field is rotated, being associated with the transformation of the fermion and the gauge field, as 
\begin{eqnarray}
\Phi(x) \to \Phi'(x)=e^{-i 2g_A \theta(x)} \Phi(x),
\end{eqnarray}
then the action becomes gauge invariant.
After $\Phi(x)$ takes the vacuum expectation value $\langle \phi \rangle =m$ and $\langle \chi \rangle=0$, then the starting action Eq.~(\ref{original action}) of this paper is reproduced.  Therefore, the starting Lagrangian can be understood as the gauge-fixed Lagrangian when the unitary gauge is taken.

It is, however, not easy to derive the effective action without fixing the gauge, that is, by including the unfixed $\phi=\sigma$ (Higgs mode) and $\chi=\pi$ (Nambu-Goldstone mode).  If it is allowed to ignore the position dependency, or to consider constant $\phi $ and $ \chi $, then we can proceed a step further.

In this case the point particle Lagrangian $\tilde{L}$ is modified by adding $\Delta \tilde{L}$,
\begin{eqnarray}
\Delta \tilde{L}= -2i \chi \gamma^{\mu} \gamma_5 \dot{x}_{\mu} -i g_A  \chi  \gamma^{\mu}\left( g_{\mu\nu} + i \sigma_{\mu\nu} \right)  x^{\lambda} F_{\lambda\nu} - \left\{ \phi^2+7 \chi^2 + 2i \gamma_5 \phi  \chi \right\}.
\end{eqnarray}
and the formula is modified to 
\begin{eqnarray}
\mathcal{L}_{\mathrm{eff}}&=&-\frac{1}{8\pi^2} \int_0^{\infty} \frac{ds}{s^3} e^{-m^2 s} \frac{(g_{+}s)^2 \mathcal{G}}{\mathrm{Im} \cosh(g_{+}Xs)} \times \frac{(g_{-}s)^2 \mathcal{G}}{\mathrm{Im} \cosh(g_{-}Xs)} \nonumber \\
&&~~~~~~~~~~~~~\times \frac{1}{2} \left\langle e^{i \int_0^s ds'~\Delta \tilde{L}(s')} \times \left( \cos \bar{\bm{X}}'_{+}(s) +\cos \bar{\bm{X}}'_{-} (s)\right) \right\rangle'_{s \to is} . 
\end{eqnarray}
However, this is not enough to discuss the ``Higgs mechanism'' properly, since the Nambu-Goldstone mode, $\partial_{\mu} \chi(x)$, has not been included.

Also, to discuss the anomaly-cancelled realistic model, extension of the H-E formula in the non-Abelian gauge theory with parity violation is required.  The non-Abelian version of the H-E formula has already been found in the vector coupling case (namely QCD) by M. J. Duff and M. Ram\'on-Medrano~\cite{non-Abelian H-E1} and by I. A. Batalin, S. G. Matinyan and G. K. Savvidy~\cite{non-Abelian H-E2}, so the extension of this paper to non-Abelian gauge theories seems to be possible, and if this can be done, then the difficulties discussed will be clarified.  We hope to tackle this problem in the near future.

Before ending the paper, we add a comment on anomaly-related fields that are actively being studied at present.  One field is the study of semimetals~\cite{semimetal1,semimetal2}, and another is the chiral magnetic effect in heavy ion collisions \cite{chiral magnetic effect1, chiral magnetic effect2} (see also reviews in \cite{chiral magnetic effect3}).  These studies are based on Weyl fermions and the chiral anomaly along the direction opened by Nielsen and Ninomiya \cite{Nielsen-Ninomiya}.

Let us first understand that the expectation value of the gauge field $\langle A_{\lambda} \rangle$ is the chemical potential for a current $J^{\lambda}(x)$ to which the gauge field couples.  For example, the usual chemical potential $\mu$, which is able to control the particle number $n(x)=\psi (x)^{\dagger} \psi(x)$, is the expectation value of the time component of a vector field, $\mu=\langle A_{0}\rangle$. Similarly, the chiral chemical potential $\mu_5$ is introduced in order to control the chiral charge, $n_5(x) =\psi (x)^{\dagger}\gamma_5 \psi(x)$, and is the expectation value of the time component of a purely axial vector field, $\mu_5=\langle A_{5, 0}\rangle$.  In a recent study of semimetals \cite{semimetal1, semimetal2}, the more general chemical potentials are introduced by $b_{\lambda}=\langle A_{5, \lambda}\rangle$. Then, $b_0$ is $\mu_5=b_0$, while $\bm{b}$ is a kind of constant magnetic field that can control the direction of the spin $\psi(x)^{\dagger} \bm{\sigma} \psi (x)$.  In the same manner, we may introduce $e_{\lambda}=\langle A_{\lambda}\rangle$, where $e_0=\mu$, and $\bm{e}$ is kind of a constant electric field that controls the momentum direction of the Fermi particle, $\bm{p}=\psi(x)^{\dagger}(\gamma_5\bm{\sigma})\psi(x)=\psi(x)^{\dagger}\bm{\alpha}\psi(x)$.  The $\bm{p}$ includes the Zitterbewegung. 

There are two cases. 

Case A: If the transition between topologically different vacua really occurs, via instantons etc., then the action has a term proportional to $\mathcal{G}=\frac{1}{4}F_{\mu\nu}\tilde{F}^{\mu\nu}=\bm{E}\cdot \bm{B}$, even without the chemical potentials $b_{\lambda}$.  The coefficient of this action is a constant and is called $\theta$.  The chiral magnetic effect in heavy ion collisions is an example of this case.

Case B: If the chemical potentials $b_{\lambda}$ are introduced, then the bias between positive and negative chiralities of fermions, or that between positive and negative spins, is realized.  Due to the chiral anomaly, or the Atiyah-Singer index theorem, these biases can be compensated by an additional term in the action that is proportional to $\mathcal{G}$.  This issue was studied in Refs.~\cite{semimetal1, semimetal2}.  However, the introduction of chemical potentials violates the Lorentz, the translational, or the rotational symmetries, as is understood from $b_{\lambda}=\langle A_{5, \lambda}\rangle \ne 0$.  Therefore, the coefficient in the action is no longer constant but is position dependent, $\theta(x) \propto b_{\lambda}x^{\lambda}$. The semimetals are examples of this case. 

Now, we will compare our calculation with the above two cases.  We have studied the ``vacuum,'' being considered an infinitely large space, where the Lorentz, the translational, and the rotational symmmetries hold.  Therefore, we have not introduced the chemical potentials, and choose $A_{\mu}=\frac{1}{2} x^{\nu} F_{\nu\mu}$, which is sufficient for our purpose to generate the constant electromagnetic fields. So, Case B has not applied to our calculation.  Also, we have not considered the topologically nontrivial configurations for gauge fields.  So, our effective action does not include a term proportional to $\mathcal{G}$.  The reason is that the transition between topologically different ``vacua'' seems difficult at zero-temperature vacuum birefringence experiment. So, Case A has also not applied to our calculation.

We have mentioned that the anomalous Ward identity or the Atiyah-Singer index theorem was proved by Schwinger in 1951, using the proper time method.  To consider this issue in our context we will introduce, in addition to the gauge field, the ``pion field" $\pi(x)$ which couples to the divergence of the current as $\pi(x)\partial J(x)$.  Then, following Schwinger, 
$\langle \partial J(x) \rangle$ can be estimated and the anomalous Ward identity can be derived. Without introduction of the ``pion field," the $\mathcal{G}$ term does not appear in $\mathcal{L}_{\mathrm{eff}}(A_{\mu}(x))$, but the effective action of a coupled system of pion and gauge fields, $\mathcal{L}_{\mathrm{eff}}(\pi(x),~A_{\mu}(x))$, can accommodate the term $\pi(x) \mathcal{G}$.

The discussion so far is the standard way to understand the vacuum.  The ``vacuum'' we are considering may not be an infinitely large space, but is a restricted asymmetric region in the laboratory, where a strong magnetic field is applied in one direction and a laser beam propagates in another direction.  In that situation the ``vacuum'' may violate the Lorentz, translational, and rotational symmetries.  Then, it may be reasonable to choose an ansatz, $A_{\lambda}(x)=a_{\lambda}+ b x_{\lambda}+\frac{1}{2} x^{\rho} F_{\lambda\rho}$, since this ansatz not only generates the constant fields $F_{\lambda\rho}$, but also gives the chemical potential $a_{\lambda}$ and the longitudinal component of the gauge field, $\partial A=b$.  The chemical potential $a_{\lambda}$ controls a combination of vector and axial currents, $g_V J^{\lambda}(x) + g_A J_5^{\lambda}(x)$.  This ansatz has a possibility of opening a new feature of the ``vacuum,'' treating it as if it were a plasma, a semimetal, or other abnormal materials.  

A problem remains, however, as to whether the bias is able to be produced in the laboratory, between positive and negative chiralities, between up and down spins, between particle and anti-particle numbers, etc.  To pursue this possibility it is important to understand a difference between the chemical potentials $(\bm{e},\bm{b})$ and the usual electric and magnetic fields $(\bm{E},\bm{B})$.  The former chemical potentials $(\bm{e},\bm{b})$ couple selectively to fermions, but the electromagnetic fields couple more generally.

\section*{Acknowledgments}
The authors thank to G.-C.~Cho, Toshiaki Kaneko, Kiyoshi Kato, and Koichi Seo for fruitful discussions.
%
%
\newpage
\renewcommand{\theequation}{A.\arabic{equation}}
\setcounter{equation}{0}
\section*{Appendix A: Diagonalization of the background field}
The matrix notation for the field strength ${(\bm{F})_{\mu}}^{\nu}={F_{\mu}}^{\nu}$, has the Lorentz vector indices, the down index $\mu$ as a row index and the upper index $\nu$ as the column index.  This form of ${(\bm{F})_{\mu}}^{\nu}$ can be multiplied naturally to form $\bm{F}^n$.  However, this real matrix is symmetric with respect to the electric field $\bm{E}$, but is antisymmetric with respect to the magnetic field $\bm{B}$.
This matrix can be diagonalized by choosing a proper ``frame" in which the matrix gives four eigenvalues written in terms of the electric fields $\bm{E}$ and the magnetic fields $\bm{B}$, or by the Lorentz-invariant quantities $\mathcal{F}$ and $\mathcal{G}$.  We will call this proper ``frame'' the ``diagonal frame.''

Let us briefly review the derivation in Ref.~\cite{Schwinger} of the four eigenvalues Eq.~(\ref{eigen-values}).  First we have to understand the following two identities:
\begin{eqnarray}
F_{\mu}^{~\lambda}\tilde{F}_{\lambda}^{~\nu}=\tilde{F}_{\mu}^{~\lambda}F_{\lambda}^{~\nu}=-\delta_{\mu}^{~\nu}~\mathcal{G}, ~\mathrm{and}~\tilde{F}_{\mu}^{~\lambda}\tilde{F}_{\lambda}^{~\nu}-F_{\mu}^{~\lambda}F_{\lambda}^{~\nu}=2~\delta_{\mu}^{~\nu}~ \mathcal{F}.  \label{two identities}
\end{eqnarray}
It is important that the right-hand sides become identity matrices.  So, if we introduce an eigenvalue $F'$ and its eigenfunction $\psi_{\mu}$ for the matrix ${F_{\mu}\;}^{\nu}$, satisfying ${F_{\mu}\;}^{\nu} \psi_{\nu}=F' \psi_{\mu}$,  then we also have the eigenvalue for $\tilde{F}_{\mu}^{~\nu}$, as $\tilde{F}_{\mu}^{~\nu} \psi_{\nu}=(\mathcal{G}/F') \psi_{\mu}$ from the first identity in Eq. (\ref{two identities}).  Now, we can estimate the eigenvalue for any product of the matrices $\bm{F}$ and $\bm{\tilde{F}}$, under a constraint coming from the second identity in Eq.~(\ref{two identities}).  This is
\begin{eqnarray}
(F')^4 + 2 \mathcal{F} (F')^2- \mathcal{G}^2=0. \label{eigen-value equation}
\end{eqnarray}
Four solutions of this algebraic equation give four eigenvalues, $F'_{(\lambda)}~(\lambda=\pm1, ~\pm2)$ given in Eq.~(\ref{eigen-values}).
The above explanation is beautiful, but it is sometimes enough to understand simply that Eq.~(\ref{eigen-value equation}) is nothing but the usual eigenvalue equation, $\det (\bm{F}-F'\bm{1} )=0$, where the matrix is $\bm{F}_{\mu}^{~\nu}$, and its roots are Eq.~(\ref{eigen-values}).

In the following, we will utilize this ``diagonal frame,'' and replace $\bm{F}$ with one of its eigenvalues $F'$. 

When we utilize the ``diagonal frame" the indices of which we label as $(\mu), (\nu), \ldots$, it is important to understand that the transformation from the original Lorentz frame to this ``diagonal frame'' is ``not an orthogonal'' but a unitary transformation.  We assume the diagonal index $(\mu)$ runs in the order of $(+1, -1, +2, -2)$.  

The diagonalization crucially depends on the matrix form of $\bm{F}_{\mu}^{~\nu}$, having the lower (covariant) index as the row number and the upper (contravariant) index as the column number.  Therefore, the diagonalization of the matrix $\bm{F}_{\mu}^{~\nu}$ differs from that of the matrix $\bm{F}_{\mu\nu}$.  The diagonalization of $\bm{F}_{\mu}^{~\nu}$ can be done Lorentz invariantly, since in this case the product of $\bm{F}_{\mu}^{~\nu}$ 's becomes a simple product without using the metric tensor.  On the other hand, the diagonalization of $\bm{F}_{\mu\nu}$ is done more smoothly in the Euclidean frame.  So, the transformation of the Lorentz frame to the diagonal frame resembles that from the Minkowski frame to  the Euclidean frame. 

If we label the four eigenvectors, corresponding to the eigenvalue $F'_{(\lambda)}~(\lambda=(+1, -1, +2, -2)$ as $\psi_{\mu}^{~(\lambda)}$, then the diagonalization of $\bm{F}$ can be performed by
\begin{eqnarray}
V^{\mu}_{~(\mu)} \bm{F}_{\mu}^{~\nu} U_{\nu}^{~(\nu)}= F'_{(\mu)} \delta_{(\mu)}^{~~(\nu)}, \label{diagonalization of F}
\end{eqnarray}
where $U$ and $V$ are defined by
\begin{eqnarray}
U_{\nu}^{~(\nu)}= \psi_{\nu}^{~(\nu)},~V^{\mu}_{~(\mu)}=g^{\mu\nu} \psi_{\nu}^{~(\nu)}g_{(\nu)(\mu)},  \label{definition of U and V}
\end{eqnarray}

We can choose the eigenvectors so that the following orthonormality, and completeness may hold:
\begin{eqnarray}
&&\bullet ~\sum_{\mu} V^{\mu}_{~(\lambda)} U_{\mu}^{~(\rho)}=\delta_{(\lambda)}^{~(\rho)} ~~\mathrm{(orthonormality ~condition)},\\
&&\bullet~ \sum_{(\lambda)} V^{\mu}_{~(\lambda)} U_{\nu}^{~(\lambda)}=\delta^{\mu}_{~\nu}~~\mathrm{(completeness ~condition)}.
\end{eqnarray}
From these, we obtain the metric in the diagonal frame.  For this purpose, we need the explicit expression of the eigenvectors $\psi_{\mu}^{~(\lambda)}$ in a generic background field of $\bm{F}$.  We choose it as $\bm{B}=(B, 0, 0)$, and $\bm{E}=(E \cos \theta, E \sin \theta)$.  Then, we have
\begin{eqnarray}
g_{(\mu)(\nu)}=V^{\mu}_{~(\mu)} ~g_{\mu\nu} ~V^{\nu}_{~(\nu)} =\pmatrix{1 && 0 && 0 && 0 \cr 
0 && 1 && 0 && 0 \cr 
0 && 0 && -1 && 0 \cr 
0 && 0 && 0 && -1}
\end{eqnarray} 
which can be inverted and gives $g_{\mu\nu}=U_{\mu}^{~(\mu)}g_{(\mu)(\nu)} U_{\nu}^{~(\nu)}$.

Now, we understand how to  manage the propagators in the ``diagonal frame'', even if they have different types of Lorentz indices before diagonalization.  The propagator $\Delta(s', s'')_{\mu}^{~\nu}$ is considered to be basic, and is diagonalized when $\bm{F}$ is diagonalized, namely  
\begin{eqnarray}
&&\Delta(s', s'')_{\mu}^{~\nu} ~V^{\mu}_{~(\mu)} U_{\nu}^{~(\nu)}=\Delta(s', s'')_{(\mu)}^{~(\nu)} \nonumber \\
&&=\pmatrix{\Delta_{(1)}  && 0 && 0 && 0 \cr 
0 && \Delta_{(-1)}  && 0 && 0 \cr 
0 && 0 &&\Delta_{(2)}  && 0 \cr
0 &&  0 && 0 && \Delta_{(-2)} }_{(\mu)}^{~(\nu)}(s', s''),
\end{eqnarray}
but $\Delta(s', s'')_{\mu\nu}$ is expressed otherwise after the diagonalization, such as 
\begin{eqnarray}
&&\Delta(s', s'')_{\mu\nu} V^{\mu}_{~(\mu)} V^{\nu}_{~(\nu)} =\Delta(s', s'') _{(\mu)}^{~~(\lambda)} g_{(\lambda)(\nu)} \nonumber \\
&&=\pmatrix{\Delta_{(1)}  && 0 && 0 && 0 \cr 
0 && \Delta_{(-1)} && 0 && 0 \cr 
0 && 0 && -\Delta_{(2)}  && 0 \cr
0 &&  0 && 0 && -\Delta_{(-2)}}_{(\mu)(\nu)}(s'-s'').
\end{eqnarray}

Therefore, we can raise and lower indices even in the diagonal frame, but as was mentioned before the best way is to keep the matrix representation just before the end, and is to diagonalize the expression at the end.
\renewcommand{\theequation}{B.\arabic{equation}}
\setcounter{equation}{0}
\section*{Appendix B: Periodic propagator in the background field}
From the definition of the propagator Eq.~(\ref{def of the propagator}), its diagonal component (eigenvalue) $\Delta(s', s'')_{(\lambda)}=\frac{1}{2i}\langle x_{(\lambda)}(s') x^{(\lambda)}(s'') \rangle$ satisfies
\begin{eqnarray}
\left\{\frac{\partial^2}{\partial s^{'2}} + 2(g_V F'_{(\lambda)})\frac{\partial}{\partial s^{'}} -2 (g_AF'_{(\lambda)})^2 \right\} \Delta(s', s'')_{(\lambda)} = \delta(s'-s''),
\end{eqnarray}
First we will obtain the propagator, denoted by $\Delta^{(0)}(s', s'')_{(\lambda)}$, in the infinite region of $-\infty < s'-s'' < +\infty$.  

After factorizing the vector-like interaction by
\begin{eqnarray}
\Delta^{(0)}(s', s'')_{(\lambda)}= e^{-g_V F_{(\lambda)}(s'-s'')} \times \Delta^{(1)}(s', s'')_{(\lambda)},
\end{eqnarray}
the latter factor $\Delta^{(1)}$ satisfies
\begin{eqnarray}
\left\{\frac{\partial^2}{\partial s^{'2}} -(g_V^2+2g_A^2)F^{'2}_{(\lambda)} \right\} \Delta^{(1)}(s', s'')_{(\lambda)} = \delta(s'-s'').
\end{eqnarray}
This one-dimensional Helmholtz equation can be easily solved by imposing the boundary condition of $\Delta^{(1)}(s', s'')_{(\lambda)} \to 0$ at $|s'-s''| \to \infty$, and hence we have the propagator in the infinite region as follows:
\begin{eqnarray}
\Delta(s', s'')^{(0)}_{(\lambda)}=-\frac{1}{2\sqrt{(g_V^2+2g_A^2)}|F'_{(\lambda)}|}  e^{-\left\{ g_V F'_{(\lambda)}(s'-s'') + \sqrt{(g_V^2+2g_A^2)}|F'_{(\lambda)}| |s'-s''| \right\}},
\end{eqnarray}
where $F'_{(\lambda)}$ is written in terms of $\mathcal{F}$ and $\mathcal{G}$ as in Eq.~(\ref{eigen-values}).  


In the above we consider the infinite region, $-\infty  < s'-s'' < \infty$.  However, the region necessary in our problem is not the infinite region, but the finite region for $-s< s'-s'' < s$.  To remedy this point, we repeat the finite regions successively and form the infinite real axis from $-\infty$ to $+\infty$ with periodic boundary conditions. (This is similar to the process to obtain the finite-temperature propagator from the zero-temperature one.) Then, the delta function should be replaced by\footnote{The delta function of the infinite space, $\delta(s'-s'')=\frac{1}{2\pi} \int_{-\infty}^{\infty} dp~ e^{-ip(s'-s'')}$ is replaced by the delta function of the finite space with periodic boundary conditions, $\delta(s'-s'')=\frac{1}{s} \sum_{n=-\infty}^{\infty} e^{-2\pi ni(s'-s'')/s}= \sum_{n=-\infty}^{\infty}\delta(s'-s''-ns)$ which is the Poisson summation formula.}
\begin{eqnarray}
\delta(s'-s'') \to \sum_{n=-\infty}^{+\infty} \delta(s'-s''-ns) 
\end{eqnarray}
Taking into account this modification, the propagator in the finite region $-s<s'-s''<s$ is obtained as follows:
\begin{eqnarray}
\Delta(s'-s'')_{(\lambda)}=\sum_{n=-\infty}^{\infty} \Delta^{(0)}(s'-s''-ns)_{(\lambda)},
\end{eqnarray}
The summation over $n$ for $s'>s''$ or $s'<s''$ yields the same expression, irrespective of the sign of $F'_{(\lambda)}$, and we have the periodic propagator Eq.~(\ref{propagator}): 
\begin{eqnarray}
\Delta(s'-s'')_{(\lambda)}&&=\frac{-1}{4(g_{+}-g_{-})F'_{(\lambda)}} \times \nonumber \\
&&\left\{ \frac{e^{-2g_{+} F'_{(\lambda)} (s'-s''-\epsilon(s'-s'')\frac{s}{2})}}{\sinh (g_{+}F'_{(\lambda)}s)}-\frac{e^{-2g_{-} F'_{(\lambda)} (s'-s''-\epsilon(s'-s'')\frac{s}{2})}}{\sinh (g_{-}F'_{(\lambda)}s)} \right\}, 
\end{eqnarray}
where $\epsilon(s'-s'')$ is a step function, and $g_{\pm}$ is given in Eq.~(\ref{g_pm}). 

As a check we can see that it reproduces the propagator without the external field \cite{Strassler}, namely
\begin{eqnarray}
\Delta(s', s'')_{(\lambda)} \vert_{F'_{(\lambda)} \to 0} = \frac{1}{2}|s'-s''| - \frac{(s'-s'')^2}{2s} + \mathrm{const}.
\end{eqnarray}

\renewcommand{\theequation}{C.\arabic{equation}}
\setcounter{equation}{0}
\section*{Appendix C: Estimation of $\langle \bar{\bm{\mathcal{F'}}}(s) \rangle' $, $\langle (\bar{\bm{\mathcal{F'}}}(s))^2\rangle'$ and $\langle (\bar{\bm{\mathcal{G'}}}(s))^2 \rangle' $}
With the help of the propagator Eq.~(\ref{propagator}) and the vertices Eqs.~(\ref{V(1)}), (\ref{V(0)}), and  (\ref{effective V(1)}), the diagrams in Fig.~3 can be estimated as follows:
\begin{eqnarray}
&&(3-1)= LP_{[1][2][3][4]}, ~(3-2)=LP_{[1][2]}~LP_{[3][4]}, ~(3-3)=LP_{[1][2][3]}\times V^{(0)}_{[4]}, \nonumber \\
&&(3-4)=LP_{[1][2]}\times V^{(0)}_{[3]}V^{(0)}_{[4]}, ~(3-5)=V^{(0)}_{[1]}V^{(0)}_{[2]}V^{(0)}_{[3]}V^{(0)}_{[4]},
\end{eqnarray}
where the loop diagram $LP_{[1][2]\ldots}$ with the number of external legs $(1, 2, \cdots)$ is defined in Eq.~(\ref{LP}).  Here $[i]$ stands for the two indices $[i]=(\mu_i\nu_i)$ following from the $i$th external leg to the loop diagram.  

What we need to estimate is $\langle \bar{\bm{\mathcal{F'}}}(s) \rangle' $, $\langle (\bar{\bm{\mathcal{F'}}}(s))^2\rangle'$, and $\langle (\bar{\bm{\mathcal{G'}}}(s))^2 \rangle' $, where the external indices of the leg are contracted with those of the other leg by $g_{\mu\nu}$ or $\epsilon_{\mu\nu\lambda\rho}$.  Taking into account the combinatorial factors properly, the definition of $\bar{\bm{\mathcal{F'}}}(s)$ and $\bar{\bm{\mathcal{G'}}}(s)$ in terms of $B_{\mu\nu}(s)$, gives the following results:
\begin{eqnarray} 
&&\langle \bar{\bm{\mathcal{F'}}}(s) \rangle' = \frac{1}{4} g^{[1][2]}(LP_{[1][2]}+V^{(0)}_{[1]}V^{(0)}_{[2]}),~ \langle\bar{\bm{\mathcal{G'}}}(s) \rangle' = \frac{1}{4} \epsilon^{[1][2]} (LP_{[1][2]}+V^{(0)}_{[1]}V^{(0)}_{[2]}), \\
&&\langle \bar{\bm{\mathcal{F'}}}(s)^2 \rangle' = \frac{1}{16} (g^{[1][2]}g^{[3][4]}+g^{[1][3]}g^{[2][4]}+g^{[1][4]}g^{[2][3]}) \times \left(LP_{[1][2][3][4]} + LP_{[1][2]}LP_{[3][4]} \right. \nonumber \\
&& \left. +\frac{4}{3} LP_{[1][2][3]}V^{(0)}_{[4]} + (LP_{[1][2]} V^{(0)}_{[3]}V^{(0)}_{[4]}+LP_{[3][4]} V^{(0)}_{[1]}V^{(0)}_{[2]}) +\frac{1}{3}V^{(0)}_{[1]}V^{(0)}_{[2]}V^{(0)}_{[3]}V^{(0)}_{[4]} \right), ~~~~ \\
&& \langle\bar{\bm{\mathcal{G'}}}(s)^2 \rangle'= \frac{1}{16} (\epsilon^{[1][2]}\epsilon^{[3][4]}+\epsilon^{[1][3]}\epsilon^{[2][4]} +\epsilon^{[1][4]}\epsilon^{[2][3]}) \times \left(LP_{[1][2][3][4]} + LP_{[1][2]}LP_{[3][4]} \right. \nonumber \\
&&\left. +\frac{4}{3} LP_{[1][2][3]}V^{(0)}_{[4]} + (LP_{[1][2]} V^{(0)}_{[3]}V^{(0)}_{[4]} +LP_{[3][4]} V^{(0)}_{[1]}V^{(0)}_{[2]})+\frac{1}{3}V^{(0)}_{[1]}V^{(0)}_{[2]}V^{(0)}_{[3]}V^{(0)}_{[4]} \right), ~~~\label{sum of graphs}
\end{eqnarray}
where $g^{[1][2]} \equiv g^{\mu_1\mu_2}g^{\nu_1\nu_2}$ and $\epsilon^{[1][2]}\equiv \frac{1}{2} \epsilon^{\mu_1\nu_1\mu_2\nu_2}$.


To obtain more explicit expressions, if we define
\begin{eqnarray}
&&V_{\mu\nu}^{(1)[\alpha, \beta]} = \epsilon_{\mu\nu}^{~~~\alpha'\beta'} \times g_A F_{\alpha'\beta'}^{~~~\alpha\beta}, \\
&&F_{\alpha'\beta'}^{~~~\alpha\beta} \equiv \frac{1}{2} ( \delta_{\alpha'}^{~\alpha} F_{\beta'}^{~\beta}+\delta_{\beta'}^{~\beta} F_{\alpha'}^{~\alpha}),
\end{eqnarray}
then we have
\begin{eqnarray}
&&\bullet~ g^{[1][2]}V_{[1]}^{(1)[\alpha_1, \beta_1]}V^{(1)}_{[2]~[\alpha_2, \beta_2]} \nonumber \\
&&~~~~~=\frac{1}{2}g_A^2 \left\{ 2F^{\alpha_1}_{~~[\alpha_2}F^{\beta_1}_{~~\beta_2]}+\delta^{\alpha_1}_{~~[\alpha_2} (\bm{F}^2)^{\beta_1}_{~~\beta_2]}+\delta^{\beta_1}_{~~[\beta_2} (\bm{F}^2)^{\alpha_1}_{~~\alpha_2]} \right\}, \\
&&\bullet~ g^{[1][2]}V_{[1]}^{(1)[\alpha_1, \beta_1]}V^{(0)}_{[2]} \nonumber \\
&&~~~~~=g_A (\delta_{\alpha_1'}^{~\alpha_1} F_{\beta_1'}^{~\beta_1} + F_{\alpha_1'}^{~\alpha_1} \delta_{\beta_1'}^{~\beta_1}  ) \times (-1) (g_V \tilde{F}^{\alpha'_1\beta'_1}+ig_A F^{\alpha'_1\beta'_1}), \\
&&\bullet~ \epsilon^{[1][2]}V_{[1]}^{(1)[\alpha_1, \beta_1]}V_{[2]}^{(1)[\alpha_2, \beta_2]} \nonumber \\
&&~~~~= -\frac{1}{2}g_A^2 \epsilon^{\alpha_1'\beta_1'\alpha_2'\beta_2'} (\delta_{\alpha_1'}^{~\alpha_1} F_{\beta_1'}^{~\beta_1} + F_{\alpha_1'}^{~\alpha_1} \delta_{\beta_1'}^{~\beta_1}  ) \times (\delta_{\alpha_2'}^{~\alpha_2} F_{\beta_2'}^{~\beta_2} +  F_{\alpha_2'}^{~\alpha_2} \delta_{\beta_2'}^{~\beta_2}), \\
&&\bullet~ \epsilon^{[1][2]}V_{[1]}^{(1)[\alpha_1, \beta_1]}V_{[2]}^{(0)} \nonumber \\
&&~~~~~= -g_A (\delta_{\alpha_1'}^{~\alpha_1} F_{\beta_1'}^{~\beta_1} + F_{\alpha_1'}^{~\alpha_1} \delta_{\beta_1'}^{~\beta_1}  ) \times (-1) (g_V F^{\alpha'_1\beta'_1}-ig_A \tilde{F}^{\alpha'_1\beta'_1}). \label{contraction of external legs}
\end{eqnarray}

As an example, we estimate $g^{[1][2]}g^{[3][4]} LP_{[1][2][3][4]}$ a piece of 
$\langle \bar{\bm{\mathcal{F'}}}(s)^2 \rangle'$.  It can be reduced to
\begin{eqnarray} 
&&g^{[1][2]}g^{[3][4]} LP_{[1][2][3][4]}  \nonumber \\
&&~=(-2i)^4 g_A^4 \times \left(g^{[1][2]}V_{[1]}^{(1)[\alpha_1, \beta_1]}V^{(1)}_{[2]~[\alpha_2, \beta_2]}g^{[3][4]}V_{[3]}^{(1)[\alpha_3, \beta_3]}V^{(1)}_{[4]~[\alpha_4, \beta_4]} \right) \nonumber \\
&&~\times \int_0^s ds_1ds_2ds_3ds_4 ~\Delta(\dot{s_1}, s_2)_{\beta_1}^{~\alpha_2} \Delta(\dot{s_2}, s_3)^{\beta_2}_{~\alpha_3} \Delta(\dot{s_3}, s_4)_{\beta_3}^{~\alpha_4} \Delta(\dot{s_4}, s_1)^{\beta_4}_{~\alpha_1}. \nonumber \\
\end{eqnarray}

To obtain the first nontrivial effective action, we need only the terms of $O(g_V \bm{F})^4$.  The prefactor existing in the product of the four vertex functions already gives $\bm{F}^4$, so that only the field-independent, free propagators contribute to the result.  This is true for all the Feynman graphs examined in~ Eq.~(\ref{sum of graphs}).
Therefore, for these fourth-order graphs, all the propagators are replaced by the free propagator:
\begin{eqnarray}
\Delta_0(\dot{s'}, s'')_{\alpha\beta}=g_{\alpha\beta}\left(\frac{1}{2} \epsilon(s'-s'')- \frac{(s'-s'')}{s} \right)=g_{\alpha\beta} \left(\frac{1}{2} \epsilon(t'-t'')- (t'-t'') \right), \nonumber \\
\end{eqnarray}
where $t'=s'/s$, and it is integrated over $0< t' <1$.  Therefore, the $s'$-integration gives $s^4$.  The the result behaves as $O(s g_V \bm{F})^4$.

To estimate $\langle \bar{\bm{\mathcal{F'}}}(s) \rangle' $, however, its prefactor comes from the two vertex functions, giving $(s g_A \bm{F})^2$, so that in this case,  the propagator should be expanded up to the second order in $(s\bm{F})^2$, and we use the following expansion:
\begin{eqnarray}
\Delta(\dot{s'}, s'')_{(\lambda)} &=& -u+\frac{1}{12} ~g_V~ (sF'_{(\lambda)})\left(-1+\delta(t'-t'')+12 u^{2}\right) \nonumber \\
&+&\frac{1}{6} \left(g_V^2 +\frac{1}{2}g_A^2 \right)(sF'_{(\lambda)})^2 \left(u-4u^{3}\right)+\cdots, ~~~~~~
\end{eqnarray}
where $u=(s'-s'')/s-\frac{1}{2} \epsilon(s'-s'')=(t'-t'')- \frac{1}{2} \epsilon(t'-t'')$, for the dimensionless parameter $t'=s'/s$.

The $\langle \bar{\bm{\mathcal{F'}}}(s) \rangle'$ is obtained on the ``diagonal frame'' as 
\begin{eqnarray}
&&\langle \bar{\bm{\mathcal{F'}}}(s) \rangle'=\frac{1}{4} g^{[1][2]}(LP_{[1][2]}+V^{(0)}_{[1]}V^{(0)}_{[2]}) \\
&&=-s^2 g_A^2~ \frac{1}{2} \sum_{(\alpha) \ne (\beta)} \left(F'_{(\alpha)} + F'_{(\beta)}\right)^2 \int_0^1 dt'~dt'' \Delta(\dot{t'}-t'')_{(\alpha)} \Delta(\dot{t''}-t')_{(\beta)} \nonumber \\
&&~~+ s^2 \left((g_V^2+g_A^2)\mathcal{F} - 2i g_Vg_A \mathcal{G}\right) \label{expression 1} \\
&&=s^2 g_A^2  \int_0^1 dt'~dt''~\left\{ 2\sum_{(\alpha)} \left(F'_{(\alpha)}\Delta(\dot{t'}-t'')_{(\alpha)} \right)^2 - \left(\sum_{(\alpha)}F'_{(\alpha)} \Delta(\dot{t'}-t'')_{(\alpha)} \right)^2 \right. \nonumber \\
&&\left. - \left(\sum_{(\alpha)}F_{(\alpha)}^{'2} \Delta(\dot{t'}-t'')_{(\alpha)} \right) \left(\sum_{(\beta)} \Delta(\dot{t'}-t'')_{(\beta)}\right) \right\} + s^2 \left((g_V^2+g_A^2)\mathcal{F} - 2i g_Vg_A \mathcal{G}\right). \label{expression 2}~~~~~~~~~
\end{eqnarray}
Using the expansion in $s$ of the propagator, we can sum over the eigenvalues, and we obtain Eq.~(\ref{mathcal F}). In this estimation the product of two delta functions appears, and we consider it to be zero for the following reason. The coefficient $\delta(0)$ of the product $\delta(t'-t'')^2=\delta(0) \delta(t'-t'')$ can be regularized, by using the zeta function $\zeta(s)=\sum_{n=1}^{\infty} n^{-s}$, as
\begin{eqnarray}
\delta(0)= \sum_{n=-\infty}^{\infty} e^{-2\pi i n \times 0} =\sum_{n=-\infty}^{\infty}~1=1+2 \zeta(0)=0.
\end{eqnarray}
This is equivalent to simply discarding the linear divergences in the ultraviolet region arising from the coincidence of the two points, $t'$ and $t''$.

Next we estimate $\langle \bar{\bm{\mathcal{F'}}}(s)^2 \rangle'$ and $\langle \bar{\bm{\mathcal{G'}}}(s)^2 \rangle'$.  In this case the $s'$-integration is carried out with the free propagator, giving one of the following:
\begin{eqnarray}
&&s^2 \int_0^1 dt_1dt_2 ~\Delta_0(\dot{t_1}-t_2)_{\beta_1\alpha_2} \Delta_0(\dot{t_2}-t_1)_{\beta_2\alpha_1}= -\frac{1}{12} s^2 g_{\beta_1\alpha_2}~g_{\beta_2\alpha_1}, \\
&&s^3 \int_0^1 dt_1dt_2dt_3 ~\Delta_0(\dot{t_1}-t_2)_{\beta_1\alpha_2} \Delta_0(\dot{t_2}-t_3)_{\beta_2\alpha_3} \Delta_0(\dot{t_3}-t_1)_{\beta_3\alpha_1}=0, \\
&&s^4 \int_0^1 dt_1dt_2dt_3 dt_4~\Delta_0(\dot{t_1}-t_2)_{\beta_1\alpha_2} \Delta_0(\dot{t_2}-t_3)_{\beta_2\alpha_3} \Delta_0(\dot{t_3}-t_4)_{\beta_3\alpha_4}\Delta_0(\dot{t_4}-t_1)_{\beta_4\alpha_1} \nonumber \\
&&=\frac{1}{720}~s^4 g_{\beta_1\alpha_2}~ g_{\beta_2\alpha_3} ~g_{\beta_3\alpha_4}~g_{\beta_4\alpha_1}.
\end{eqnarray}
Therefore, the triangle loop graphs vanish as expected.

Then, we have
\begin{eqnarray}
&&LP_{[1][2]}=\frac{1}{3}~s^2~V_{[1]}^{(1)[\alpha_1\alpha_2]}V^{(1)}_{[2][\alpha_2\alpha_1]}, \\
&&LP_{[1][2][3]}=0, \\
&&LP_{[1][2][3][4]}=\frac{1}{45}~s^4~V_{[1]}^{(1)[\alpha_1\alpha_2]}V^{(1)}_{[2][\alpha_2\alpha_3]}V_{[3]}^{(1)[\alpha_3\alpha_4]}V^{(1)}_{[4][\alpha_4\alpha_1]}.
\end{eqnarray}

Using these expressions and Eq.~(\ref{contraction of external legs}), which give the way to contract external legs in Eq.~(\ref{sum of graphs}), we obtain the expressions of the individual graphs:
\begin{eqnarray}
&&\diamond~ (g^{[1][2]}g^{[3][4]}+g^{[1][3]}g^{[2][4]}+g^{[1][4]}g^{[2][3]}) LP_{[1][2][3][4]} \nonumber \\
&&=\frac{8}{45}~ s^4 g_A^4  \left( 5 \mathcal{F}^2 + ~\mathcal{G}^2 \right), \nonumber \\
&&\diamond~ (g^{[1][2]}g^{[3][4]}+g^{[1][3]}g^{[2][4]}+g^{[1][4]}g^{[2][3]}) LP_{[1][2]}LP_{[3][4]} \nonumber \\
&&=\frac{32}{9}~ s^4 g_A^4 \left( 3\mathcal{F}^2 -\mathcal{G}^2 \right),  \\
&&\diamond~ (g^{[1][2]}g^{[3][4]}+g^{[1][3]}g^{[2][4]}+g^{[1][4]}g^{[2][3]}) (LP_{[1][2]}V^{(0)}_{[3]}V^{(0)}_{[4]}+ LP_{[3][4]}V^{(0)}_{[1]}V^{(0)}_{[2]})\nonumber \\
&&=\frac{64}{3}~s^4 g_A^2 \left\{ (g_V^2+g_A^2) \mathcal{F}^2 -2i ~g_V g_A \mathcal{F} \mathcal{G} \right\}, \\
&&\diamond~(g^{[1][2]}g^{[3][4]}) V^{(0)}_{[1]} V^{(0)}_{[2]} V^{(0)}_{[3]}V^{(0)}_{[4]} \nonumber \\
&&~~=16 ~s^4 \left\{ (g_V^2+g_A^2) \mathcal{F} - 2i~g_Vg_A ~\mathcal{G}  \right\}^2, \\
&&\diamond~(\epsilon^{[1][2]}\epsilon^{[3][4]}+\epsilon^{[1][3]}\epsilon^{[2][4]}+\epsilon^{[1][4]}\epsilon^{[2][3]}) LP_{[1][2][3][4]} \nonumber \\
&&~~=\frac{8}{45} ~s^4 g_A^4 \left( \mathcal{F}^2 +5\mathcal{G}^2 \right), \\
&&\diamond (\epsilon^{[1][2]}\epsilon^{[3][4]}+\epsilon^{[1][3]}\epsilon^{[2][4]}+\epsilon^{[1][4]}\epsilon^{[2][3]}) LP_{[1][2]}LP_{[3][4]} \nonumber \\
&&~~=\frac{32}{9} s^4 g_A^4 ( -\mathcal{F}^2 +3\mathcal{G}^2 ), \nonumber \\
&&\diamond~(\epsilon^{[1][2]}\epsilon^{[3][4]}+\epsilon^{[1][3]}\epsilon^{[2][4]}+\epsilon^{[1][4]}\epsilon^{[2][3]}) (LP_{[1][2]}V^{(0)}_{[3]}V^{(0)}_{[4]}+ LP_{[3][4]}V^{(0)}_{[1]}V^{(0)}_{[2]}\nonumber \\
&&~~=\frac{64}{3} s^4 g_A^2 \left( (g_V^2+g_A^2) \mathcal{G}^2 + 2i~g_Vg_A ~\mathcal{F} \mathcal{G} \right), \\
&&\diamond~(\epsilon^{[1][2]}\epsilon^{[3][4]}) V^{(0)}_{[1]} V^{(0)}_{[2]} V^{(0)}_{[3]}V^{(0)}_{[4]} \nonumber \\
&&~~=16 ~s^4 \left\{ (g_V^2+g_A^2) \mathcal{G} + 2i~g_Vg_A ~\mathcal{F}  \right\}^2.
\end{eqnarray}
In estimating these, we have used $\mathrm{tr} (\bm{F}^2)=-4\mathcal{F}$, $\mathrm{tr} (\bm{F}^4)=8 \mathcal{F}^2+4 \mathcal{G}^2$, and $\det(\bm{F}_{\mu}^{~\nu})=-\mathcal{G}^2$.

Now, using the above expressions for individual graphs in Eq.~(\ref{sum of graphs}), we obtain Eqs.~(\ref{mathcal F squared}) and (\ref{mathcal G squared}).

\renewcommand{\theequation}{D.\arabic{equation}}
\setcounter{equation}{0}
\section*{Appendix D: Proof of the Jacobian $\mathcal{J}=1$}
In this appendix we show that the Jacobian $\mathcal{J}$ associated with the change of fermionic variable $\psi'(x)=i\gamma^5 \psi(x)$ and $\overline{\psi}'(x)=\overline{\psi}(x)i\gamma^5$ is 1.  Noting that the fermionic Jacobian is the inverse of the bosonic case, we have
\begin{eqnarray}
\mathcal{J}^{-1}= \mathrm{Det}_{\{x', \;x''\}} \langle x' | i \gamma_5 | x'' \rangle,
\end{eqnarray} 
where $\mathrm{``Det"}$ consists of the determinant $\det'$ with respect to the two end points of the path $P=P(x', ~x'')$ and the determinant $\det''$ for the Dirac spinors (or for the spin degrees of freedom).

Again following Schwinger's proper time method \cite{Schwinger}, we can estimate $\mathcal{J}^{-1}$ using the results so far obtained.  That is, we will evaluate
\begin{eqnarray}
\mathcal{J}^{-1}= \lim_{i s \to 0} {\det}' \left\{ \langle x'(s) | x''(0)  \rangle' \times {\det}^{''} \left\langle i \gamma_5 \times e^{+i\frac{1}{2}\bar{B}_{\mu\nu}(s) \sigma^{\mu\nu}} \right\rangle' \right\}.  \label{Jacobian}
\end{eqnarray}

Both the determinant with respect to the spin ${\det}^{''}$ and the determinant ${\det}^{'}$ with respect to the final and initial points $(x'', x')$ give 1, but in order to prove this, we have to be careful about the different initial and final points in the evaluation of the Jacobian.  

So, we will consider the mode expansion, not of the closed string so far discussed, but of the open string along the path $P(x', ~x'')$: 
\begin{eqnarray}
&&P(x', ~x''):~x^{\mu}(s')=x^{\mu}_0 (s') + \frac{1}{\sqrt{s}} \sum_{n=1}^{\infty} \left(a^{\mu}_n e^{-2\pi n i (s'/s)}+a^{\mu \dagger}_n e^{2\pi ni(s'/s)} \right) \nonumber \\
&&\equiv x_0^{\mu} (s')+ x_q^{\mu}(s'), \label{mode expansion of the open string}
\end{eqnarray}
where $x^{\mu}_0(s')$ is a classical solution of the point particle theory satisfying $x^{\mu}_0(s)=x'^{\mu}$ and $x^{\mu}_0(0)=x''^{\mu}$.

The equation of motion of the point particle described by the Lagrangian (\ref{Lagrangian of point particle}) is
\begin{eqnarray}
\ddot{\bm{x}}+ 2 (g_V \bm{F}) \dot{\bm{x}} -2 (g_A \bm{F})^2 \bm{x}=0,
\end{eqnarray}
where the matrix notation is used for the four Lorentz indices.  This is the harmonic oscillator with a friction, so that the solution can be sought in the form $e^{\bm{\Omega}s}$, and the two special solutions for $\bm{\Omega}_{\pm}=-2g_{\pm} \bm{F}$ give
\begin{eqnarray}
\bm{x}(s')=e^{-2 g_{+} \bm{F} s'}\bm{x}_{+} + e^{-2 g_{-} \bm{F} s'}\bm{x}_{-},
\end{eqnarray}
with two arbitrary constant vectors $(\bm{x}_{+}, ~\bm{x}_{-})$ which are to be determined by the starting and ending positions, $(x', ~x'')$.  For this purpose, the diagonalization of the matrix $\bm{F}$ is necessary, but it is familiar to us now.  In the diagonalized frame, the four components are denoted by $(\lambda)=(\pm 1, ~\pm 2)$, and 
\begin{eqnarray}
&&x_{+(\lambda)}=\frac{x'_{(\lambda)}- e^{-2g_{-}F'_{(\lambda)} s} x''_{(\lambda)}}{e^{-2g_{+}F'_{(\lambda)} s}-e^{-2g_{-}F'_{(\lambda)} s}},  \\
&&x_{-(\lambda)}=-\frac{x'_{(\lambda)}- e^{-2g_{+}F'_{(\lambda)} s} x''_{(\lambda)}}{e^{-2g_{+}F'_{(\lambda)} s}-e^{-2g_{-}F'_{(\lambda)} s}},
\end{eqnarray}
where $F'_{(\lambda)}$ are given in Eq.~(\ref{eigen-values}), and so the classical solution $x_0^{\mu}(s)$ is determined.

First we estimate the trace for the spin degrees of freedom, {\it i.e.} the second factor in Eq.~(\ref{Jacobian}),
\begin{eqnarray}
[\mathrm{\mathrm{det}''~ of ~the~ spin}] = \lim_{i s \to 0} {\det}^{''} \left\langle i \gamma_5 \times e^{+i\frac{1}{2} \sigma^{\mu\nu} \bar{B}_{\mu\nu}(s)} \right\rangle'  \label{the second factor of Jacobian}
\end{eqnarray}

Here, we have to estimate the expectation value when the classical solution exists, so we have to separate the classical solution as in Eq.~(\ref{mode expansion of the open string}), $x_0^{\mu}=x_0^{\mu} (s')+ x_q^{\mu}(s')$, and take the expectation value for the remaining  ``quantum part,'' namely
\begin{eqnarray}
\langle x_q^{\alpha}(s') x_q^{\beta}(s'') \rangle'=2i~\Delta^{\alpha\beta}(s', s'').
\end{eqnarray}

When we take the expectation value, the field $x^{\mu}(s')$ is replaced by its classical solution $x^{\mu}_0(s')$, or is contracted with the other $x^{\nu}(s'')$ and gives the propagator $\Delta^{\mu\nu}(s', s'')$.  In the former case, the original dimension of the length, $[L]$, possessed by $x^{\mu}(s')$ will be carried by the dimension $[L]$ of the initial and final positions, $x^{' \mu}$ and $x^{'' \mu}$.  On the other hand in the latter case, the dimension of $x^{\mu}(s')$ will be finally (after integration) carried by $\sqrt{s}$.  Let's explain this a little more when $s \to 0$.

In the former case, when $s \to 0$,
\begin{eqnarray}
x_{\mu}(s')_0 =  x''_\mu+ \frac{s'}{s} (x'-x'')_\mu+\cdots, ~~\dot{x}_{\mu}(s')_0 = \frac{(x'-x'')_\mu}{s} +\cdots.
\end{eqnarray}
After integration,
\begin{eqnarray}
\bar{B}_{\mu\nu}(s) = \langle \bar{B}_{\mu\nu}(s)\rangle_0 \equiv \frac{1}{4} g_A \epsilon_{\mu\nu\beta\gamma}F^{\alpha\beta}(x'+x'')_\alpha(x'-x'')_\beta
\end{eqnarray}
is constant when $s \to 0$. So, the dimension $[L]^2$ is carried by $x^{' \mu}$ and $x^{'' \mu}$.

In the latter case, the propagator becomes, when $s \to 0$,
\begin{eqnarray}
\Delta(\dot{s'}, s'')_\lambda &=& -\left( \frac{s'-s''}{s}- \epsilon(s'-s'') \right) \nonumber \\
&+& (g_+ +g_-) (F'_\lambda s) \left( \frac{1}{6} + \left(\frac{s'-s''}{s}\right)^2-\frac{|s'-s''|}{s} \right) \\
&= & f_1(t) + (F'_\lambda s) f_2(t), 
\end{eqnarray}
where $t$ is the dimensionless parameter $t=s'/s$.

Whenever the number of the contraction and the replacement to the propagator is increased by one from the former replacement with the classical solution, $s$ appears, since the variable of the propagator $s'-s''$ is always integrated from $0$ to $s$,
\begin{eqnarray}
&&\int_0^s ds_1 \int_0^s ds_2 \cdots \left\{ \Delta(\dot{s}_1-s_2) \Delta(\dot{s}_2-s_3) \cdots \right\}= \left(s \int_0^1 dt _1\right) \left(s \int_0^1 dt _2 \right) \nonumber \\
&&\times \left\{ \left(f_1+ (F'_\lambda s)  f_2 \right)(t_1-t_2) \left(f_1+ (F'_\lambda s)  f_2 \right)(t_2-t_3)  \cdots \right\}.
\end{eqnarray}
Therefore, the terms contracted with the propagators are higher order in $s$, and hence in the limit of $s \to 0$, only the classical part remains, that is 
\begin{eqnarray}
[\mathrm{\mathrm{det}''~ of ~the~ spin}] = {\det}^{''}\left( i \gamma_5 \times e^{+i\frac{1}{2} \sigma^{\mu\nu} \langle \bar{B}_{\mu\nu}(s) \rangle_0 } \right),
\end{eqnarray} 
which can be estimated as
\begin{eqnarray}
[\mathrm{\mathrm{det}''~ of ~the~ spin}]=e^{\mathrm{tr}' \left(  i \frac{\pi}{2} \gamma_5 + i \frac{1}{2} \sigma_{\mu\nu}\langle \bar{B}_{\mu\nu}(s) \rangle_0 \right) }=1.
\end{eqnarray} 

Therefore, we have  
\begin{eqnarray}
\mathcal{J}^{-1}= \lim_{i s \to 0}{\det_{\{x', \;x''\}}}'\langle x'(s) | x''(0) \rangle'
\end{eqnarray}

Here, the transition amplitude $\langle x'(s) | x''(0) \rangle'$ becomes
\begin{eqnarray}
\langle x'(s) | x''(0) \rangle' &\propto& e^{i S_0(s;~x',~x'')}\times  \prod_{\lambda=\pm1, \pm2} \prod_{n=1}^{\infty} \frac{1}{\ell(n, \lambda)} \\
&\propto& e^{i S_0(s;~x',~x'')} \times \frac{(g_{+}s)^2 \mathcal{G}}{\mathrm{Im} \cosh(g_{+}Xs)} \times \frac{(g_{-}s)^2 \mathcal{G}}{\mathrm{Im} \cosh(g_{-}Xs)}.
\end{eqnarray}

The contribution of the action integral from the classical solution appears in the present case in that the initial and final points differ, and it reads\begin{eqnarray}
S_0(s;~x', ~x'')=-\frac{1}{2} \sqrt{g_V^2-g_A^2} \sum_{\lambda=\pm1,~\pm2} x_{+\lambda} F'_{\lambda} \left( e^{2\sqrt{g_V^2-g_A^2}F'_{\lambda} s} -1 \right) x_{-\lambda}.
\end{eqnarray}
Now, $\langle x'(s) | x''(0) \rangle'$ is explicitly given, the $(x', x'')$-dependent factor comes from the classical action integral, while the $(x', x'')$-independent factor comes from the mode sum, and hence its expansion in terms of s reads
\begin{eqnarray}
\langle x'(s) | x''(0) \rangle' = e^{i\left( A(x', ~x'') \frac{1}{s} + B(x', ~x'')s + \cdots \right)} \times (1+ C s^2 + \cdots), 
\end{eqnarray}
where  the expansion in $s$ appears as the expansion in $s^2$.  This is because the expansion in $s$ appears always, associated with the external field $\bm{F}$ as the dimensionless combination $s\bm{F}$, and Lorentz invariance requires that the expansion be in $s^2(\bm{F})^2$.  Here, $A(x', ~x'')=\frac{1}{4}(x'-x'')^2$ and is independent of the external field $\bm{F}$.  The determinant with external field is always evaluated as the ratio to the free Jacobian, $\frac{\mathcal{J}|_{F \ne 0}}{\mathcal{J}|_{F=0}}$, and so the determinant to be evaluated is
\begin{eqnarray}
\lim_{is \to 0} \det' \frac{\langle x'(s) | x''(0) \rangle' |_{F \ne 0}}{\langle x'(s) | x''(0) \rangle' |_{F =0}} = \lim_{is \to 0} e^{\mathrm{Tr} \ln \left(\frac{\langle x'(s) | x''(0) \rangle' |_{F \ne 0}}{\langle x'(s) | x''(0) \rangle' |_{F = 0}}\right)}=1.
\end{eqnarray}

Now, $\mathcal{J}=1$ has been proved.


\end{document}